\documentclass[12pt]{article}
\usepackage{latexsym}
\usepackage{amsfonts}
\usepackage{epsfig}

\catcode`\@=11

\def\section{\@startsection{section}{1}{\z@}{3.5ex plus 1ex minus
 .2ex}{2.3ex plus .2ex}{\bf}}

\def\thesubsection{\arabic{section}.\arabic{subsection}}
\renewcommand{\subsection}[1]{\addtocounter{subsection}{1}
\vspace{2.5mm}\par\noindent {\it \thesubsection . #1}\par
 \vspace{0.5mm} }
\catcode`\@=12

\mathchardef\varGamma="0100
\mathchardef\varDelta="0101
\mathchardef\varTheta="0102
\mathchardef\varLambda="0103
\mathchardef\varXi="0104
\mathchardef\varPi="0105
\mathchardef\varSigma="0106
\mathchardef\varUpsilon="0107
\mathchardef\varPhi="0108
\mathchardef\varPsi="0109
\mathchardef\varOmega="010A

\def\bfone{\relax{\rm 1\kern-.35em 1}}

\DeclareFontFamily{U}{rsf}{}
\DeclareFontShape{U}{rsf}{m}{n}{
  <5> <6> rsfs5 <7> <8> <9> rsfs7 <10-> rsfs10}{}
\DeclareMathAlphabet\Scr{U}{rsf}{m}{n}

\begin{document}

\begin{titlepage}

\begin{flushright}
\hskip 5.5cm \hskip 1.5cm
\vbox{\hbox{CERN-TH/2003--136}
\hbox{SPIN-03/19}
\hbox{ITP-03/31}
\hbox{hep-th/0306185}} 
\end{flushright} 

\vskip 2cm
\centerline{{\large\bf New $D=4$ gauged supergravities from}}
\vskip 0.35cm 
\centerline{{\large\bf ${\Scr N}=4$
orientifolds with fluxes}}
\vskip 1cm
\centerline{Carlo Angelantonj${}^{1,\dagger}$, Sergio 
Ferrara${}^{1,2,\ddagger}$ and 
Mario Trigiante${}^{3,\flat}$}
\vskip 0.5cm
\centerline{\small\it ${}^1$ 
CERN, Theory Division, CH 1211 Geneva 23, Switzerland}

\centerline{\small\it ${}^2$ INFN, Laboratori Nazionali di Frascati, Italy}

\centerline{\small\it ${}^3$ Spinoza Institute, Leuvenlaan 4 NL-3508, Utrecht,
The Netherlands}

\centerline{\small e-mail: $^{\star}$ carlo.angelantonj@cern.ch , $^{\ast}$
sergio.ferrara@cern.ch, $^\flat$
M.Trigiante@phys.uu.nl} 

\vskip 1cm
\begin{abstract}
We consider classes of $T_6$--orientifolds, where the orientifold projection 
contains an inversion $I_{9-p}$ on $9-p$ coordinates, transverse 
to a ${\rm D}p$--brane.
In absence of fluxes, the massless sector of these models corresponds to 
diverse  forms of ${\Scr N}=4$ supergravity, with six bulk vector 
multiplets coupled to ${\Scr N}=4$ Yang--Mills theory on the branes. 
They all differ in the choice of the duality symmetry corresponding to 
different embeddings of $SU(1,1)\times SO(6,6+n)$ in $Sp(24+2n,\mathbb{R})$, 
the latter being the full group of duality rotations. Hence, these 
Lagrangians are not related by local field redefinitions. 
When fluxes are turned on one can construct new gaugings of 
${\Scr N}=4$ supergravity, where the twelve bulk vectors gauge some 
nilpotent algebra which, in turn, depends on the choice of fluxes.  
\end{abstract}

\end{titlepage}

\section{Introduction}

New string or M--theory models are obtained turning on $n$--form fluxes, 
which allow, in general, the lifting of vacua, supersymmetry breaking and 
moduli stabilisation \cite{ps}--\cite{Tripathy:2003qw}. 
Examples of such new solutions are IIB and IIA orientifolds
\cite{cargese,orient,horava,pol,revs}, where the orientifold 
projection (in absence of fluxes) preserves ${\Scr N}=4$ or 
${\Scr N}=2$ supersymmetries.

Recently, the $T_6/\mathbb{Z}_2$ orientifold with ${\Scr N}=4$ 
supersymmetry \cite{fp,kst} and $K_3\times T_2/\mathbb{Z}_2$ orientifold 
\cite{Tripathy:2003qw} with ${\Scr N}=2$ supersymmetry have been the
subject of an extensive study. In these cases, turning on NS--NS and 
R--R three--form fluxes allows to obtain new string vacua with vanishing 
vacuum energy, reduced supersymmetry and moduli stabilisation
\cite{gkp,fp,kst,Kachru:2002sk,Tripathy:2003qw}. 
These features can all be understood in terms of an effective 
gauged supergravity, where certain axion symmetries are gauged
\cite{D'Auria:2002tc,D'Auria:2002th,D'Auria:2003jk}. These are 
generalised no--scale models \cite{Cremmer:1983bf,Ellis:1983sf}.

In the present investigation, we consider more general four--dimensional 
orientifolds with fluxes (both in type IIB and IIA) where the orientifold 
projection involves an inversion $I_{9-p}$ on $9-p$ coordinates, 
transverse to the ${\rm D}p$--brane world--volume, thus generalising 
the $T_6/\mathbb{Z}_2$ orientifold (with $p=3$) constructed by 
Frey--Polchinski \cite{fp} and Kachru--Shulz--Trivedi \cite{kst} 
(see also \cite{Berg:2003ri} for a derivation of the complete low--energy 
supergravity from T--dialysed Type I theory in ten dimensions). 
Interestingly, their low--energy descriptions are all given in terms 
of ${\Scr N}=4$ supergravity with six vector supermultiplets from
the closed--string sector, coupled to an ${\Scr N}=4$ Yang--Mills 
theory living on the ${\rm D}p$--brane world--volume. 

However, despite the uniqueness of ${\Scr N}=4$ supersymmetry, the
low--energy actions crucially differ in the choice of the manifest ``duality 
symmetries'' of the Lagrangian, since different sets of fields survive
the orientifold projection, and therefore different symmetries are
manifestly preserved. Leaving the brane degrees of freedom aside, these 
duality symmetries are specified by their action on the (twelve) bulk 
vectors. Actually, ${\Scr N}=4$ supergravity demands that such 
symmetries be contained in $SU(1,1)\times SO(6,6)$ \cite{sugra41,sugra42}
and act on the vector 
field strengths and their duals as symplectic  $Sp(24,\mathbb{R})$ 
transformations \cite{Gaillard:1981rj}. 
On the other hand, the symmetries 
of the Lagrangian correspond to block--lower--triangular symplectic matrices, 
whose block--diagonal components have a definite action on the vector 
potentials \cite{Andrianopoli:2002aq,Dabholkar:2002sy,deWit:2002vt}.
For instance, in the orientifold models containing an $I_{9-p}$ inversion, the 
block--diagonal symmetries always include $GL(9-p,\mathbb{R})\times 
GL(p-3,\mathbb{R})$, as maximal symmetry of the $GL(6,\mathbb{R})$ 
associated to the moduli space of the six--torus metrics.
The lower--triangular block contains the axion symmetries of the R--R 
scalars and of the NS--NS ones originating from the $B$--field, whenever
present\footnote{For example, the latter is not 
present in the $p=3$ case, i.e. the $T_6/\mathbb{Z}_2$ orientifold.}.

In the sequel, we describe all nilpotent algebras $N_p$ 
\cite{Andrianopoli:1996bq}, corresponding to axion 
symmetries of the R--R and NS--NS scalars for all orientifold models.
All $N_p$'s are nilpotent subalgebras of $so(6,6)$, are generically
non--abelian and contain central charges. There are four of 
them in type IIB ($p=3,\,5,\,7,\,9$) with dimensions $15,\,23,\,23,\,15$ 
respectively, while there are only three of them in type IIA ($p=4,\,6,\,8 $) 
of dimensions $20,\,24,\,20$, respectively. A common feature of these algebras 
is that they always contain fifteen R--R axionic symmetries, while 
the extra symmetries correspond to NS--NS $B$--field axions in 
the bi--fundamental of $GL(9-p,\mathbb{R})\times GL(p-3,\mathbb{R})$.

A further R--R axion symmetry originates from the $SU(1,1)$, which acts as 
electric--magnetic duality on the gauge fields living on the brane 
world--volume. The corresponding axion field can be identified with the
$C_{p-3}$ R--R field, as dictated by the coupling
\begin{equation}
\int_{\varSigma_{p+1}} \, C_{p-3}\wedge F\wedge F \,, \label{wzdb}
\end{equation} 
where $F$ is the two--form field strength of gauge fields living on the branes.

Turning on fluxes in the orientifold models (three- and five--form fluxes in 
type IIB, two- and four--form fluxes in IIA) corresponds to 
a ``gauging'' in the corresponding supergravity Lagrangian, whose couplings 
are dictated by the particular choice of fluxes. Non--abelian gaugings 
may also occur corresponding to subalgebras of $N_p$, or quotient 
algebras $N_p/Z$, where $Z$ are some of the central generators of $N_p$. 

As an illustrative example, let us consider the $p=7$ type IIB orientifold 
defined in section 2, where the non--vanishing NS--NS and R--R fluxes are 
$H_{aij},\,F_{aij},\,G^i=\epsilon^{ab}\,\epsilon^{ijkl}\,G_{abjkl}$ 
($a,b=5,6$ and $i,j=1,\dots, 4$), and let us look at terms involving the 
axions coming from the $B$ and four--form fields, $B_{ia}$ and 
$C_{ijab}=C_{ij}\epsilon_{ab}$. Inspection of the three--form  kinetic term 
reveals a non--abelian gauge coupling proportional to 
\begin{equation}
\sqrt{-g}\, H_{aij}\, H_{\mu \nu b}\, g^{ab}\,g^{i\mu}\,g^{j\nu} \,,
\end{equation} 
as well as axion gauge couplings proportional to
\begin{equation}
\sqrt{-g}\, H_{aij}\, H_{\mu b\ell}\, g^{ab}\,g^{i\mu}\,g^{j\ell} \,,
\end{equation} 
together with similar expressions for the $F$--three form. Such terms come 
also from the reduction of type IIB four--form field. In addition, when a 
five--form flux $G^i$ is turned on an axion gauge coupling emerges of the type
\begin{equation}
\partial_\mu C_{ij}+\epsilon_{ijk\ell}\, G^k\, G^\ell_\mu \,.
\end{equation} 
where $G^\ell_\mu\,=\,g^{\ell i}\, g_{i\mu}$ are the Kaluza--Klein vectors.  
We report here only a preliminary analysis of the deformation of the 
${\Scr N}=4$ supergravity due to these new gaugings.

In the present paper we do not address either the question of unbroken 
supersymmetries or the question of moduli stabilisation, which would require 
the knowledge of the scalar potential and a study of the fermionic sector. 
However, we can anticipate that certain moduli are indeed stabilised in 
all these models, since a Higgs effect is taking place as suggested by the 
presence of charged axion couplings. 

The paper is organised as follows: in section 2, we review the 
four-dimensional $T_6 /\mathbb{Z}_2$ orientifold models, their spectra and 
their allowed fluxes. In section 3, the ${\Scr N}=4$ supergravity 
interpretation is given for the ungauged case (absence of fluxes)
and the duality symmetries exposed. The $N_p$ algebras are exhibited as 
well as their action on the vector fields. In section 4, we give a preliminary
description of gauged supergravity, for the particular case of type IIB 
orientifolds with some three--form fluxes turned on. In section 5 some 
conclusions are drawn. Finally, in appendix some useful formulae needed
to compute the quadratic part of the vector field strengths in the Lagrangian,
are given.

\section{${\Scr N}=4$ orientifolds: spectra and fluxes}

In this section we review the construction of orientifold models preserving 
${\Scr N} =4$ supersymmetries in $D=4$ \cite{revs}. This is the simplest 
setting for orientifold constructions, and consists of modding out 
type II superstrings by the world--sheet parity $\varOmega$ \cite{cargese}.
Following \cite{pol,revs}, the orientifold projection 
can be given a suggestive geometrical interpretation
in terms of non--dynamical defects, the orientifold ${\Scr O}$--planes, 
that reflect the left--handed and right--handed modes of the closed string.
Actually, one can combine world--sheet parity with other (geometrical)
operations. In general, this can affect the nature of the orientifold
planes, that, in the simplest instance of a bare $\varOmega$ have 
negative tension and R--R charge, and are $(9+1)$--dimensional (${\Scr O}9$ 
planes) since they have to respect the full Lorentz symmetry preserved by
$\varOmega$. In the present paper, we are interested in the class of models
generated by the $\varOmega I_{9-p}$ generator, where $I_{9-p}$ denotes 
the inversion on $9-p$ coordinates. Of course, $\Omega I_{9-p}$ must 
be a symmetry of the parent theory, and this is the case of type 
IIB for $p$ odd, and of type IIA for $p$ even. Actually, $\varOmega I_{9-p}$ 
reflects the action of T-duality in orientifold models. Indeed, T-duality 
itself can be thought of as a chiral parity transformation
\begin{equation}
X_{\rm L} \to X_{\rm L} \,, \qquad X_{\rm R} \to - X_{\rm R} \,,
\end{equation}
and conjugates $\varOmega$ so to get
\begin{equation}
{\Scr T}_{9-p} \varOmega {\Scr T}_{9-p}^{-1} = \varOmega I_{9-p}\,.
\end{equation}
As a result, the full ten--dimensional Lorentz symmetry is now broken
to the subgroup $SO(1,p) \times SO(9-p)$, and the closed--string sector 
involves ${\Scr O}_{9-p}$ planes sitting at the fixed points of the orbifold 
$T_{9-p}/I_{9-p}$. The associated open-string sector will then correspond to 
open strings with Dirichlet boundary conditions along $T_{9-p}$, i.e. open 
strings ending on D$(9-p)$ branes. As usual, tadpole conditions will fix the 
rank of the Chan--Paton gauge group, i.e. the total number of D-branes. 
In the present paper, however, we shall not be concerned with open--string 
degrees of freedom and we shall concentrate our analysis solely on the 
closed-string degrees of freedom. 

Before we turn to the description of specific models, a general comment is 
in order. An important requirement in the construction is that the orientifold 
group be $\mathbb{Z}_2$, i.e. its generator $\varOmega I_{9-p}$ must square 
to the identity. Although $\varOmega$ has always $\pm 1$ eigenvalues, and 
thus $\varOmega ^2 =1$, this is not the case for $I_{9-p}$. 
For example, for $p=7$ $I_2$ would correspond to a $\pi$ rotation on a
two--plane and, although its action on the bosonic degrees of freedom
is real and assigns to them a plus or minus sign according to the number of 
indices along the two--plane, its eigenvalue on spinors is 
$e^{i\pi \varSigma}$, where $\varSigma = \pm {1\over 2}$ are the two 
helicities. Thus, it does not square to the identity, but rather to
$(-1)^F$, with $F$ the (total) space--time fermion number. Therefore, 
in this case the orientifold projection needs be modified by the inclusion 
of $(-1)^{F_{\rm L}}$, with $F_{\rm L}$ the left-handed space--time fermion 
number \cite{sen}. We are thus dealing with the four-dimensional orientifolds
\begin{equation}
\left( T_{p-3} \times T_{9-p} \right) / \varOmega I_{9-p} \left[ 
(-1)^{F_{\rm L}}\right]^{\left[{9-p \over 2}\right]} \,,
\end{equation}
where $\left[{9-p \over 2}\right]$ denotes the integer part of $(9-p)/2$. Here
we have decomposed the six-torus as
\begin{equation}
T_6 = T_{p-3} \times T_{9-p} \,,
\end{equation}
since $I_{9-p}$ only acts on the coordinates of $T_{9-p}$, while leaves
invariant those along $T_{p-3}$. As we shall see, this is a natural 
decomposition since, in the orientifold, we are left with the perturbative 
symmetry $GL(p-3) \times GL(9-p)$ of the compactification torus. To fix the 
notation, in this paper we shall label coordinates on the $T_6$ with a pair 
of indices $(i,a)$, where $i=1,\dots , p-3$ counts the coordinates not affected
by the space parity (those coordinates that would be longitudinal to the
branes), while $a=1,\ldots , 9-p$ runs over the coordinates of $T_{9-p}$
(orthogonal to the branes). As usual, Greek indices $\mu ,\nu ,\ldots$ will
label coordinates on the four--dimensional Minkowski space--time.

At this point, it is better to consider the cases $p$ odd or $p$ 
even separately. In the first case, $\varOmega I_{9-p} \left[ (-1)^{F_{\rm L}}
\right]^{\left[{9-p \over 2}\right]}$ is a symmetry in type IIB, while in
the latter case it is properly defined within type IIA.

\subsection{IIB orientifolds}

In type IIB superstring we have to consider four cases, corresponding to 
the allowed choices $p=9,7,5,3$. The massless ten-dimensional fields have
a well defined parity with respect to $\varOmega$:
\begin{eqnarray}
{\rm even:} & & G_{MN} \,, \quad\, \phi \,, \quad C_{MN} \,,
\\
{\rm odd:} & & B_{MN}\,, \quad C \,, \quad C^{(+)}_{MNPQ} \,,
\end{eqnarray}
where $G_{MN}$ is the metric tensor, $\phi$ the dilaton, $B_{MN}$ the 
Kalb--Ramond two-form, and $C_{p+1}$ are the R--R 
$(p+1)$-forms\footnote{Actually, the four-form $C_{4}^{(+)}$ is constrained
to have a self--dual field strength, a peculiarity of type IIB}. Henceforth,
it is straightforward to select the four-dimensional excitations that survive
the orientifold projection. In fact, after splitting the ten-dimensional
index $M$ in the triple $(\mu, i, a)$ labelling ${\Scr M}_{1,3} \times
T_{p-3} \times T_{9-p}$, it is evident that 
the fields with an odd (even) number of $a$--type indices are odd (even)
under the action of $I_{9-p}$. On the other hand, when present, 
$(-1)^{F_{\rm L}}$ assigns a plus sign to the NS-NS states (which originate 
from the decomposition of the product of two bosonic representations of 
$SO(8)$) and a minus sign to the R--R states (which originate from the 
decomposition of the product of two spinorial representations of $SO(8)$).
At the end, aside from the four--dimensional metric tensor, one is left with 
the massless (bosonic) degrees of freedom listed in table 1.

\begin{table}[ht]
\vskip 0.5 cm
\caption{Massless degrees of freedom for the IIB orientifolds}
\vskip 0.5 cm
\begin{center}
\begin{tabular}{|r|c|c|}
\hline
$p$ & scalars & vectors \\
\hline
9 & $g_{ij}$, $\phi$, $C_{\mu\nu}$, $C_{ij}$ & $G^{i}_{\mu}$, $C_{i\mu }$ \\
\hline
7 & $g_{ij}$, $g_{ab}$, $\phi$, $B_{ia}$, $C$, $C_{ia}$, $C_{ijkl}$, $C_{ijab}$
& $G^{i}_{\mu}$, $B_{a\mu }$, $C_{a\mu}$, $C_{ijk\mu}$ \\
\hline
5 & $g_{ij}$, $g_{ab}$, $\phi$, $B_{ia}$, $C_{\mu\nu}$, $C_{ij}$, $C_{ab}$,
$C_{iabc}$ & $G^{i}_{\mu}$, $B_{a\mu}$, $C_{i\mu }$, $C_{abc\mu}$ \\
\hline
3 & $g_{ab}$, $\phi$, $C$, $C_{abcd}$ & $B_{a\mu}$, $C_{a\mu }$ \\
\hline
\end{tabular}
\end{center}
\vskip 0.5 cm
\end{table}

However, in orientifold models it happens often that fields which are odd 
under the projection can be consistently assigned with a (quantised)
background value for the fields themselves, or for their field 
strengths. For example, in the $p=7$ case the NS--NS fields $B_{ij}$ and 
the R--R fields $C_{ij}$ are both odd with respect to the orientifold 
projection and, thus, their quantum excitations are projected out. However, 
acting on them with a $\partial_a$ derivative changes their parity, and thus 
(quantised) fluxes along the internal directions, $H_{aij}$ and $F_{aij}$, 
can be incorporated in the model. Repeating a similar analysis for the other 
cases yields the allowed fluxes listed in table 2.

\begin{table}[ht]
\vskip 0.5 cm
\caption{Allowed fluxes for the IIB 
orientifolds. $F$, $H$ and $G$ fluxes are associated to the $B$, $C_2$ and 
$C_4$ fields}
\vskip 0.5 cm
\begin{center}
\begin{tabular}{|r|c|c|}
\hline
$p$ & fluxes \\
\hline
9 & none \\
\hline
7 & $H_{ija}$, $F_{ija}$, $G_{ijkab}$
\\
\hline
5 & $H_{abc}$, $F_{iab}$, $H_{ija}$, $G_{ijabc}$
\\
\hline
3 & $H_{abc}$, $F_{abc}$ 
\\
\hline
\end{tabular}
\end{center}
\vskip 0.5 cm
\end{table}

\subsection{IIA orientifolds}

Type IIA superstring selects $p$ even, and thus leaves us with the three
cases $p=8,6,4$. Although a bare $\varOmega$ is not a symmetry in type IIA, we 
can nevertheless assign a well defined parity to the massless ten-dimensional
degrees of freedom:
\begin{eqnarray}
{\rm even:} & & G_{MN} \,, \quad\, \phi \,, \quad C_{M} \,,
\\
{\rm odd:} & & B_{MN}\,, \quad C_{MNP} \,.
\end{eqnarray}
As before, $G_{MN}$ is the metric tensor, $\phi$ the dilaton, $B_{MN}$ the
Kalb--Ramond two-form, while in this case the R--R potentials $C_{p+1}$ 
carry an odd number of indices. The additional action of $I_{9-p}$ and,
eventually, of $(-1)^{F_{\rm L}}$ thus yields the massless degrees of 
freedom listed in table 3.

Also in this case one can allow for (quantised) fluxes along the 
compactification torus, as summarised in table 4.

\begin{table}[ht]
\vskip 0.5 cm
\caption{Massless degrees of freedom for the IIA 
orientifolds}
\vskip 0.5 cm
\begin{center}
\begin{tabular}{|r|c|c|}
\hline
$p$ & scalars & vectors \\
\hline
8 & $g_{ij}$, $g_{99}$, $\phi$, $B_{i9}$, $C_{i}$, $C_{9\mu\nu}$, $C_{ij9}$ & 
$G^{i}_{\mu}$, $C_{\mu}$, $C_{i9\mu}$, $B_{9\mu}$ 
\\
\hline
6 & $g_{ij}$, $g_{ab}$, $\phi$, $B_{ia}$, $C_{a}$, $C_{i\mu\nu}$, $C_{ijk}$,
$C_{iab}$
& $G^{i}_{\mu}$, $B_{a\mu }$, $C_{ij\mu}$, $C_{ab\mu}$ \\
\hline
4 & $g_{44}$, $g_{ab}$, $\phi$, $B_{4a}$, $C_{4}$, $C_{a\mu\nu}$, $C_{abc}$
& $G^{4}_{\mu}$, $B_{a\mu}$, $C_{\mu }$, $C_{4a\mu}$ \\
\hline
\end{tabular}
\end{center}
\vskip 0.5 cm
\end{table}

\begin{table}[ht]
\vskip 0.5 cm
\caption{Allowed fluxes for the IIA 
orientifolds. $F$, $H$ and $G$ fluxes are associated to the $B$, $C_1$ and 
$C_3$ fields}
\vskip 0.5 cm
\begin{center}
\begin{tabular}{|r|c|c|}
\hline
$p$ & fluxes \\
\hline
8 & $H_{ij9}$, $G_{ijk9}$
\\
\hline
6 & $H_{aij}$, $H_{abc}$, $F_{ia}$, $G_{ijab}$
\\
\hline
4 & $H_{abc}$, $F_{ab}$, $G_{4abc}$ 
\\
\hline
\end{tabular}
\end{center}
\vskip 0.5 cm
\end{table}

\section{${\Scr N}=4$ supergravity interpretation of $T_6$ orientifolds: 
manifest duality transformations and Peccei--Quinn symmetries.}

The four--dimensional low--energy supergravities of ${\Scr N}=4$ orientifolds 
(in the absence of fluxes) can be consistently constructed as truncations of 
the unique four--dimensional ${\Scr N} =8$ supergravity which describes the
low--energy limit of dimensionally reduced type II superstrings.
Its duality symmetry group $E_{7(7)}$ acts
non linearly on the $70$ scalar fields, and linearly, as a $Sp(56,\mathbb{R})$ 
symplectic transformation, on the  $28$ electric 
field strengths and their magnetic dual. In this framework an intrinsic 
group--theoretical characterisation of the ten--dimensional origin of 
the four--dimensional fields is indeed achieved. In the so--called solvable 
Lie algebra representation of the
scalar sector \cite{Andrianopoli:1996bq, Andrianopoli:1996zg}, the scalar 
manifold 
\begin{equation}
{\Scr M}_{\rm scal} = \exp{(Solv(e_{7(7)}))} 
\end{equation}
is expressed as the group manifold generated 
by the solvable Lie algebra $Solv(e_{7(7)})$ defined through the Iwasawa 
decomposition of the $e_{7(7)}$ algebra:
\begin{equation}
e_{7(7)} = su(8)+Solv(e_{7(7)}) \,.
\end{equation}
In this framework, there is a natural one--to--one correspondence
between the scalar fields and the generators of $Solv(e_{7(7)})$.
The latter consists of the $7$ generators $H_p$ of the $e_{7(7)}$
Cartan subalgebra, parametrised by the $T_6$ radii
$R_n =e^{\sigma_n}$ together with the dilaton $\phi$, and of the shift
generators corresponding to the $63$  positive roots $\alpha$ of  $e_{7(7)}$,
which are in one--to--one correspondence with the axionic scalars
that parametrise them. This correspondence between Cartan generators and 
positive roots
on one side and scalar fields on the other, can be pinpointed by 
decomposing $Solv(e_{7(7)})$ with respect to some
relevant groups. For instance, the duality group of maximal supergravity in 
$D$ dimensions is $E_{11-D(11-D)}$ and therefore, in the solvable Lie algebra 
formalism, the scalar fields in the $D$--dimensional theory are parameters of 
$Solv(e_{11-D(11-D)})$. Since $e_{11-D(11-D)}\subset e_{7(7)}$, decomposing 
$Solv(e_{7(7)})$ with respect to $Solv(e_{11-D(11-D)})$ it is possible to 
characterise the higher--dimensional origin of the four--dimensional scalars. 
Moreover, in four dimensions the group $SL(2,\mathbb{R})\times SO(6,6)_T
\subset E_{7(7)}$, $SO(6,6)_T$ being the isometry group of the $T_6$ 
moduli--space, acts transitively on the scalars originating from 
ten--dimensional 
NS--NS fields of type II theories. These scalars therefore parametrise 
$Solv(sl(2,\mathbb{R})+ so(6,6)_T)$. Henceforth, decomposing $Solv(e_{7(7)})$ 
with respect to $Solv(sl(2,\mathbb{R})+ so(6,6)_T)$ one can achieve an 
intrinsic characterisation of the NS--NS or R--R ten--dimensional origin of 
the four--dimensional scalar fields, the R--R scalars (and the corresponding 
solvable generators) transforming in the spinorial representation of 
$SO(6,6)_T$. Finally, depending on whether we interpret the four--dimensional 
maximal supergravity as tied to type II supergravities on $T_6$ or $D=11$ 
supergravity on  $T_7$, the metric moduli are acted on transitively by 
$GL(6,\mathbb{R})_g$ or $GL(7,\mathbb{R})_g$ subgroups of $E_{7(7)}$, 
respectively. Therefore, in the two cases the metric moduli 
parametrise $Solv(gl(6,\mathbb{R})_g)$ or $Solv(gl(7,\mathbb{R})_g)$ and thus, 
decomposing $Solv(e_{7(7)})$ with respect to these two solvable subalgebras, 
depending on the higher--dimensional interpretation of the four--dimensional 
theory, we may split the axions into metric moduli of the internal torus
and into scalars deriving from dimensional reductions of ten- or 
eleven--dimensional tensor fields. The latter will parametrise nilpotent 
generators transforming in the corresponding tensor representations with 
respect to the adjoint action of $GL(6,\mathbb{R})_g$ or $GL(7,\mathbb{R})_g$.
As a result of the above decompositions, we are able to
characterise unambiguously each parameter of $Solv(e_{7(7)})$ as a
dimensionally reduced field. Let us consider the dimensional reduction of 
type II supergravities. As far as the axionic scalars are
concerned the correspondence with  roots can be summarised in
terms of an orthonormal basis $\{\epsilon_p\}$ of $\mathbb{R}^7$ 
\footnote{Now and henceforth we shall  always label by 
$n,\,m=1,\dots, 6$ the $T_6$ directions, by $i,\,j=1,\dots, p-3$ the 
directions of $T_{p-3}$ which are longitudinal to the ${\rm D}p$--brane and 
by $a,\,b=p-2,\dots , 9-p$  the directions of the transverse $T_{9-p}$. 
The four--dimensional  space--time directions are generically denoted by 
Greek letters. 
}:
\begin{eqnarray}
C_{n_1n_2\dots n_k} &\leftrightarrow & a +\epsilon_{n_1}+\dots \epsilon_{n_k} 
\,,
\label{eq1}
\\
C_{ n_1n_2\dots n_k\mu\nu} &\leftrightarrow & a +\epsilon_{m_1}+\dots 
\epsilon_{m_{6-k}}\,,\,\,\,\,\,\,(\epsilon^{n_1\dots n_k m_1\dots m_{6-k}}
\neq 0) \,,
\label{eq2}
\\
B_{n m} &\leftrightarrow & \epsilon_{n}+\epsilon_{m} \,,
\label{eq3}
\\
B_{\mu\nu} &\leftrightarrow & \sqrt{2}\,\epsilon_7 \,,
\label{eq4}
\\
G_{n m}&\leftrightarrow &\epsilon_{n}-\epsilon_{m}\,,\,\,\,\,\,\,(n\neq m) \,,
\label{eq5}
\end{eqnarray}
where
\begin{equation}
a = -{\textstyle\frac{1}{2}}
\sum_{n=1}^6\, \epsilon_n+{\textstyle\frac{1}{\sqrt{2}}}\,\epsilon_7 \,.
\end{equation}
In our notation, the $so(6,6)_T$ roots have the form $\{\pm \epsilon_n
\pm\epsilon_m\}$, where $1\le n\,<\,m\le 6$. Notice indeed that the nilpotent 
generators corresponding to non--metric axions transform in tensor 
representations of $GL(6,\mathbb{R})_g$, and this, in turn, defines the 
$GL(6,\mathbb{R})_g$ representation of the corresponding scalar. For instance, 
the $C_{n_1\dots n_k}$ parametrises the generator $T^{n_1\dots n_k}\,
=\,E_{ a +\epsilon_{n_1}+\dots \epsilon_{n_k}}$ whose transformation property 
under $GL(6,\mathbb{R})_g$ is
\begin{equation}
{\bf g}\,\in\,GL(6,\mathbb{R})_g: \quad {\bf g}\cdot T^{n_1\dots n_k}\cdot 
{\bf g}^{-1}\,=\,g^{n_1}{}_{m_1}\cdots g^{n_k}{}_{m_k}\,T^{m_1\dots m_k} \,.
\end{equation}
The roots corresponding to R--R fields are spinorial with respect to
$SO(6,6)_T$ and, depending on whether the number of their indices is even or 
odd, they belong to the root system of two $e_{7(7)}$ algebras which are
mapped into each other by the $SO(6,6)_T$ outer automorphism
(T--duality) \cite{Lu:1996ge, Bertolini:1999uz}.
These two systems naturally correspond to the reduction of IIB and IIA 
superstrings, that are indeed related by T-dualities. Hence, the $T_6$ metric 
moduli in the type IIA or B descriptions, are acted upon transitively by two 
inequivalent  $GL(6,\mathbb{R})_g$ subgroups of $E_{7(7)}$: in the former case 
$GL(6,\mathbb{R})_g$ is contained in $SL(8,\mathbb{R})\,\subset\,
E_{7(7)}$, while in the latter case $GL(6,\mathbb{R})_g$  is contained in the 
maximal subgroup $SL(3,\mathbb{R})\times SL(6,\mathbb{R})_g$ of $E_{7(7)}$.
As far as the R--R scalars are concerned, the two representations differ in 
the $SO(6,6)_T$ chirality of the 32 spinorial positive roots
\begin{eqnarray}
{\rm IIA} &:&\,{\bf 32}^-\,=\,\{{\textstyle\frac{1}{2}}
(\stackrel{\mbox{odd $+$}}
{\overbrace{\pm \epsilon_1\dots \pm \epsilon_6}})+
{\textstyle\frac{1}{\sqrt{2}}}\,
\epsilon_7\} \,,
\nonumber
\\ 
{\rm IIB} &:&\,{\bf 32}^+\,=\,\{{\textstyle\frac{1}{2}}
(\stackrel{\mbox{even $+$}}
{\overbrace{\pm \epsilon_1\dots \pm \epsilon_6}})+
{\textstyle\frac{1}{\sqrt{2}}}\,
\epsilon_7\} \,.
\end{eqnarray}
Similarly, vector potentials, and their corresponding duals, are in 
one--to--one correspondence with weights $W$ of the ${\bf 56}$ of
$E_{7(7)}$ in the two representations discussed above:
\begin{eqnarray}
C_{n_1\dots n_k\mu } &\leftrightarrow & w +\epsilon_{n_1}+\dots 
\epsilon_{n_{k}} \,,
\nonumber 
\\
B_{ m\nu} &\leftrightarrow& \epsilon_{n}-{\textstyle\frac{1}{\sqrt{2}}}
\epsilon_7\,,
\nonumber
\\
G^n_{ \mu} &\leftrightarrow& -\epsilon_{n}-{\textstyle\frac{1}{\sqrt{2}}} 
\epsilon_7\,,
\nonumber
\end{eqnarray}
where
\begin{equation}
w \,=\, -{\textstyle\frac{1}{2}}\sum_{n=1}^6\,\epsilon_n \,.
\end{equation}
The dual potentials correspond to the opposite weights $-W$.

 The above axion--root ($\varPhi\,\leftrightarrow\,\alpha$) and
vector--weight ($A_\mu\,\leftrightarrow\,W$) correspondences can
be retrieved also from inspection of the scalar and vector kinetic
terms in the dimensionally reduced type IIA or type IIB
Lagrangians \cite{Lu:1996ge, Cremmer:1997ct, Cremmer:1998px} on a 
straight torus, which have the form:
\begin{eqnarray}
\mbox{dilatonic scalars:} && -\partial_\mu \vec{h}\cdot  
\partial^\mu \vec{h} \,,\nonumber
\\
\mbox{axionic scalars:} && -
{\textstyle \frac{1}{2}}\,e^{-2\,\alpha\cdot h}\,\left(\partial_\mu 
\varPhi\cdot  \partial^\mu \varPhi\right) \,,\nonumber
\\
\mbox{vector fields:} &&-
{\textstyle\frac{1}{4}}
\, e^{-2\, W\cdot h}\,F_{\mu\nu}\, F^{\mu\nu} \,,\nonumber
\end{eqnarray}
where
\begin{equation}
\vec{h}\,=\,\sum_{n=1}^6\sigma_n\,(\epsilon_n+
{\textstyle\frac{1}{\sqrt{2}}} \epsilon_7 )-
{\textstyle\frac{1}{2}}\,\phi\,a \,,\label{hdef}
\end{equation}
and, as usual, $F_{\mu\nu} = \partial_\mu A_\nu-\partial_\nu A_\mu$.

A generic axion $\varPhi$ and its dilatonic partner $e^{\alpha\cdot
h}$ can be thought of as the real and imaginary parts of a complex
field $z$ spanning an $SL(2,\mathbb{R})/SO(2)$ submanifold, where
the $SL(2,\mathbb{R})$ group is defined by the root $\alpha$.
In the models describing type II strings on $T_{p-3}\times T_{9-p}$
orientifolds, the real part of the complex scalar $z$ spanning the
$SL(2,\mathbb{R})/SO(2)$ factor in the scalar manifold is $
C_{i_1\dots i_{p-3}}$, where $i_1,\dots i_k$ label the directions
of $T_{p-3}$, as dictated by the coupling in eq. (\ref{wzdb}).
From eqs. (\ref{eq1}) and (\ref{hdef}) one can then 
verify that ${\rm Im}(z)\,=\,e^{\alpha\cdot h}={\rm Vol}_{p-3}\,
e^{\frac{p-7}{4}\,\phi}$, where ${\rm Vol}_{p-3}$ denotes the
volume of $T_{p-3}$. The scalar ${\rm Im}(z)$ defines the
effective four--dimensional coupling constant of the super Yang--Mills 
theory on
${\rm D}p$--branes through the relation:
\begin{equation}
\frac{1}{g_{\rm YM}^2} = V_{p-3}\, e^{\frac{p-7}{4}\,\phi} \,.
\label{outscal}
\end{equation}

\begin{figure}[htbp]
\centering \epsfig{file=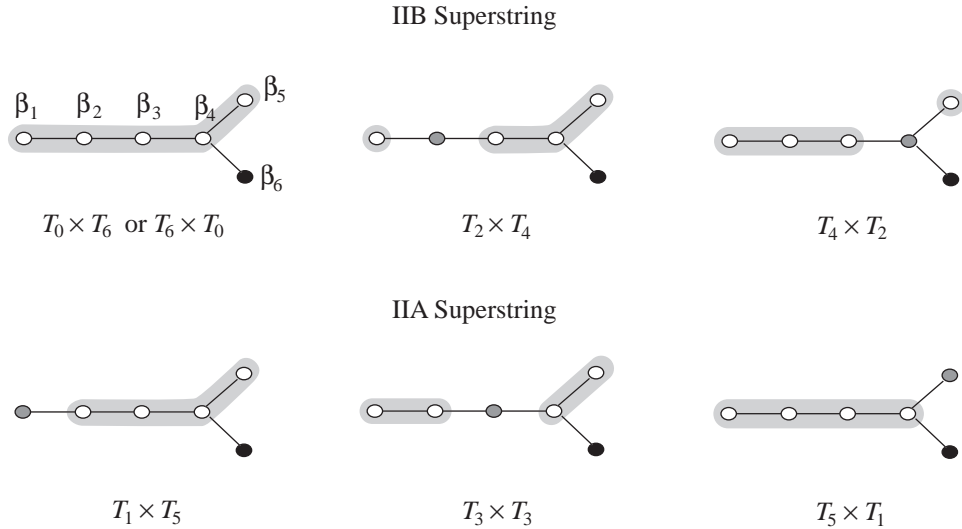} 
\caption{$SO(6,6)$ Dynkin diagrams for the $T_{p-3}\times T_{9-p}$ models. 
The shaded subdiagrams define the groups $SL(p-3,
\mathbb{R})\times SL(9-p,\mathbb{R})$ acting transitively on the metric 
moduli. The empty circles define simple roots corresponding to the metric 
moduli $g_{ij},\,g_{ab}$, the grey circle denotes a simple root corresponding 
to a Kalb--Ramond field $B_{ia}$ and the black circle corresponds to a R--R 
axion.}\label{o66}
\end{figure}

The embedding of the ${\Scr N}=4$ orientifold models $T_{p-3}\times T_{9-p}$ 
(in absence of fluxes) inside the ${\Scr N}=8$ theory (in its type IIA or 
IIB versions) is defined by specifying the embedding of the ${\Scr N}=4$ 
duality group $SL(2,\mathbb{R})\times SO(6,6)$ inside the ${\Scr N}=8$
$E_{7(7)}$ one. As far as the scalar sector is concerned, this embedding is 
fixed by the following group requirement:
\begin{equation}
SO(6,6)\,\cap\,  GL(6,\mathbb{R})_g = O(1,1)\times SL(p-3,\mathbb{R})
\times  SL(9-p,\mathbb{R}) \,.
\label{inter}
\end{equation}
 Condition (\ref{inter}) fixes the 
ten--dimensional interpretation of the fields  in the ungauged ${\Scr N}=4$ 
models (except for the cases $p=3$ and $p=9$) which, for a given $p$, is indeed
 consistent with the bosonic spectrum resulting from the orientifold reductions
listed in the previous section. In the $p=3$ and $p=9$ cases, the two
embeddings are characterised by a different interpretation of the 
scalar fields, consistent with the $T_6/\mathbb{Z}_2$ orientifold reduction 
in the presence of $D3$ or $D9$ branes. We shall denote these two 
models by $T_0\times T_6$ and $T_6\times T_0$, respectively. In these cases,
equation (\ref{inter}) in the solvable Lie algebra language amounts to 
requiring that metric moduli are related either to the $T_{p-3}$ metric 
$g_{ij}$ or to the $T_{9-p}$ metric $g_{ab}$. The scalar 
field parameterising the Cartan generator of the external $SL(2,\mathbb{R})$ 
factor is given in eq. (\ref{outscal}), while the metric modulus corresponding 
to the $O(1,1)$ in eq. (\ref{inter}) is (modulo an overall power)
\begin{equation}
O(1,1) \leftrightarrow (V_{p-3})^{9-p}\,(V_{9-p})^{11-p} 
\,.
\label{onlymet}
\end{equation}
The  axions not related to the $T_6$ metric moduli 
consist of $ C_{i_1\dots i_{p-3}}$ in the 
external $SL(2,\mathbb{R})/SO(2)$ factor, $(p-3)\,(9-p)$ moduli $B_{i a}$ in 
the bifundamental of $SL(p-3,\mathbb{R})\times  SL(9-p,\mathbb{R})$ and $15$ 
R--R moduli which we shall generically denote by $C_I$ and which span
the maximal abelian ideal $\{T^I\}$ of $Solv(so(6,6))$. The scalars $B_{ia}$ 
and $C_I$ parametrise a $15+(p-3)\,(9-p)$ dimensional subalgebra $N_p$ of 
$Solv(so(6,6))$ consisting of nilpotent generators only. In figure \ref{o66} 
the $so(6,6)$ Dynkin diagrams for the various models and the corresponding 
intersections with $gl(6,\mathbb{R})_g$, represented  by $sl(p-3,\mathbb{R})+
sl(9-p,\mathbb{R})$ subdiagrams, are illustrated. As far as the scalar fields 
are concerned, the $T_{p-3}\times T_{9-p}$ models within the same type IIA or 
IIB framework are mapped into each other by $so(6,6)_T$ Weyl transformations, 
which can be interpreted as T--dualities on an even number of directions of 
$T_6$.

We con now turn to the detailed analysis of each (IIB or IIA) orientifold 
model.

\subsection{$T_4\times T_2$ IIB orientifold with D7 branes. 
The ungauged version.}

\noindent{\it Solvable algebra of global symmetries.\ \ }
The following model (with $p=7$) describes the bulk sector of IIB superstring 
compactified on a $(T_4\times T_2)/\mathbb{Z}_2$ orientifold with D7 
branes wrapped on the $T_4$.

To this end, we describe the embedding of the scalar sector of the 
corresponding ${\Scr N}=4$ model within the ${\Scr N}=8$ by expressing 
the $so(6,6)$ Dynkin diagram $\{\beta_n\}$ in terms of the simple roots 
of $e_{7(7)}$ \footnote{In our 
conventions $\beta_1$ is the end root of the long leg and $\beta_5,\,\beta_6$ 
the symmetric roots}
\begin{eqnarray} 
\beta_1 &=& \epsilon_1-\epsilon_2\,, 
\nonumber 
\\
\beta_2 &=& \epsilon_2-\epsilon_3 \,,
\nonumber
\\
\beta_3 &=& \epsilon_3-\epsilon_4\,, 
\nonumber
\\
\beta_4 &=&\epsilon_4+\epsilon_5 \,,
\nonumber
\\
\beta_5 &=& -\epsilon_5+\epsilon_6\,, 
\nonumber
\\
\beta_6 &=&  
-{\textstyle \frac{1}{2}}
(\sum_{n=1}^6 \epsilon_n)+{\textstyle\frac{1}{\sqrt{2}}} \epsilon_7 = a \,.
\nonumber
\end{eqnarray}
According to eq. (\ref{eq1}), the root $\beta_6$ corresponds to the
ten--dimensional R--R scalar $C_{0}$, and thus identifies
the type IIB duality group $SL(2,\mathbb{R})_{\rm IIB}$.
The Dynkin diagram of the external $SL(2,\mathbb{R})$ factor in the 
isometry group consists, instead, of the single root
\begin{equation}
\beta = a+\epsilon_1+\epsilon_2+\epsilon_3+\epsilon_4 \,.
\end{equation}
It is useful to classify the positive roots according to their grading with 
respect to three relevant $O(1,1)$ groups generated by the Cartan operators 
$H_{\beta},\,H_{\lambda^4},\,H_{\lambda^6}$ and parametrised by the  moduli
$\beta\cdot h,\,h_4,\,h_6$:
\begin{eqnarray}
O(1,1)_0 &\rightarrow &  e^{\beta\cdot h}\,=\,V_4 \,,
\nonumber
\\
O(1,1)_1&\rightarrow& e^{h_4}\,=\,(V_4)^{\frac{1}{4}}\,(V_2)^{\frac{1}{2}} \,,
\nonumber
\\
O(1,1)_2&\rightarrow &e^{h_6}\,=\,e^{-\phi} \,,
\end{eqnarray}
where we have denoted by $\lambda^n$ the $so(6,6)$ simple weights,
$\lambda^n\cdot \beta_m\,=\,\delta^n_m $. $O(1,1)_0$ is generated by 
the Cartan generator of the external $SL(2,\mathbb{R})$ and 
$O(1,1)_1,\,O(1,1)_2$ are 
in $GL(4,\mathbb{R})\times GL(2,\mathbb{R})$, the former corresponding to 
the metric modulus given in eq. (\ref{onlymet}). In table 5 we list 
the axionic fields of the model together with the corresponding generator of 
$Solv(sl(2,\mathbb{R}))+Solv(so(6,6))$, for each of which the $O(1,1)^3$ 
grading and the $SL(4,\mathbb{R})\times SL(2,\mathbb{R})$ 
representations are specified. The indices 
$i,\,j$ and $a,\,b$ label as usual the directions of the torus which are 
longitudinal ($T_4$) and transverse $(T_2)$ to the D--branes.

\begin{table}[ht]
\vskip 0.5 cm 
\caption{Axionic fields for the $T_4\times T_2$ IIB orientifold, generators
of $Solv(so(6,6))$, $O(1,1)^3$ gradings, and $SL(4, \mathbb{R}) \times 
SL(2,\mathbb{R})$ representations.}
\vskip 0.5 cm
\begin{center}
{\small \begin{tabular}{|c|c|c|c|c|}
\hline
$GL(4)\times GL(2)$--rep. & generator & root & field & dim. 
\\\hline
--- & $T_{(0,0,0)}$ & $\{\epsilon_i-\epsilon_j,\,\epsilon_a-\epsilon_b\}$ 
{\scriptsize ($i<j,\,a>b$)} & $\{g_{ij},\,g_{ab}\}$ & 7 
\\\hline
${\bf (1,1)}_{(0,0,1)}$ & $T_0$ &$a $ & $C_0$ & 1  
\\\hline
${\bf (4,2)}_{(0,1,0)}$  & $T^{1ia}$ & $\epsilon_i+\epsilon_a $& $B_{ia}$ & 8 
\\\hline
${\bf (4,2)}_{(0,1,1)}$ & $T^{2ia}$ &$a +\epsilon_i+\epsilon_a$ & $C_{ia}$ & 8 
\\\hline
${\bf (6,1)}_{(0,2,1)}$ & $T^{ij}$ & $a+(\epsilon_i+\epsilon_a)+(\epsilon_j+
\epsilon_b)$ & $C_{ij\,ab}\,\equiv\, C_{ij}\,\epsilon_{ab}$ & 6 
\\\hline\hline
${\bf (1,1)}_{(2,0,0)}$ & $T$ & $\beta=a+\epsilon_1+\epsilon_2+\epsilon_3+
\epsilon_4 $ &$C_{ijkl}\,\equiv\,c$ & 1 
\\\hline
\end{tabular}}
\end{center}
\vskip 0.5 cm
\end{table}

The fields $B_{ia}$ and $C_{ia}$ transform in the representation ${\bf (4,4)}$ 
of $SL(4,\mathbb{R})\times SO(2,2) $ where $SO(2,2)=SL(2,\mathbb{R})\times 
SL(2,\mathbb{R})_{\rm IIB}$, and therefore will be collectively denoted by   
$\varPhi^{\lambda}_i$, where $\lambda =(\alpha,\,a)=1,\,2,\,3,\,4$ labels the 
${\bf 4}$ of $SO(2,2)$, with a choice of basis corresponding to the 
invariant metric $\eta_{\lambda\sigma}\,=\,{\rm diag}(+1,\,+1,\,-1,\,-1)$. Its 
expression in terms of the fields $B_{ia}$ and $C_{ia}$ is
\begin{equation}
\varPhi^\lambda_i = {\textstyle\frac{1}{\sqrt{2}}}
\{C_{i2}-B_{i1},\,B_{i2}+C_{i1},\,
B_{i1}+C_{i2},\,-B_{i2}+C_{i1}\} \,.
\end{equation}
We shall use the same notation for the corresponding generators, $\{T^i_\lambda
\}\equiv\{T^{1ia},\,T^{2ia}\}$.

From the assigned gradings one can conclude that the generators $T_0$,
$T^i_\lambda$ and $T^{ij}$ close a 23--dimensional nilpotent solvable 
subalgebra $N_7$ of $Solv(so(6,6))$.
The non--trivial commutation relations are determined by the grading and the 
index structure of the generators, and read
\begin{eqnarray}
\left[T_{0},\,T_{\lambda}^i\right]&=& M_{\lambda }{}^{\lambda^\prime }\,
T_{\lambda^\prime}^i \,,
\nonumber
\\
\left[T_{\lambda}^i,\,T_{\lambda^\prime}^j\right]&=&\eta_{\lambda\lambda
^\prime}\,\,T^{ij} \,. \label{commutator}
\end{eqnarray}
 where $M_{\lambda }{}^{\lambda^\prime }$ is a nilpotent generator acting on 
the ${\bf 4}$ of $SO(2,2)$ 
 which, for our choice of basis, can be cast in the form
\begin{equation}
M_{\lambda }{}^{\lambda^\prime } = 
{\textstyle\frac{1}{2}}\,\left(
\matrix{0 & -1 & 0 & -1
\cr 
1 & 0 & 1 & 0
\cr 
0 & 1 & 0 & 1
\cr 
-1 & 0 & -1 & 0}\right) \,.
\end{equation}

\noindent{\it Infinitesimal transformations.\ \ } Let us consider 
now the infinitesimal transformations of the scalar fields generated by
$T_{0}$,  $T_{\lambda i}$ and $T_{ij}$. For simplicity 
we shall restrict our analysis to those points in the moduli space where
the only non-vanishing scalars are $\varPhi^\lambda_i$, $C^{ij}$ and $C$.
The corresponding coset representative thus takes the simple form
\begin{equation}
L = \exp{\left(C_{ij}\,T^{ij}\right)}\,
\exp{\left(\varPhi^{\lambda}_i\,T_{\lambda }^i\right)}\,
\exp{\left(C\,T_{0}\right)} \,,
\end{equation}
and its associated left--invariant one--form is
\begin{eqnarray}
L^{-1} d L &=& (L^{-1}\partial_{0} L)\, d C+(L^{-1}\partial_{\lambda
}^i L)\, d\varPhi^{\lambda }_i+(L^{-1}\partial^{ij} L)\,
dC_{ij}
\nonumber
\\
&=& T_0\, dC+d\varPhi^{\lambda}_i\,(\delta_\lambda^{\lambda^\prime}- C
M_\lambda{}^{\lambda^\prime})\, T_{\lambda^\prime
}^i+{\textstyle\frac{1}{2}}\,T^{ij}\,d\varPhi^{\lambda }_i \varPhi_{\lambda
j}+T^{ij}\, dC_{ij} \,.
\end{eqnarray}
In general, the action of an element $T_\varLambda$ on the coset representative
can be expressed as:
\begin{equation}
L^{-1} T_\varLambda L = k_\varLambda^\alpha L^{-1} \partial_\alpha L \,,
\end{equation}
where the $k_\varLambda$ are the corresponding Killing vectors. In the case
at hand, from eq. (\ref{commutator}), we can derive
\begin{eqnarray}
L^{-1} T_{0} L &=& T_0+ \varPhi^{\lambda}_i\,M_\lambda{}^{\lambda^\prime}
\,T_{\lambda^\prime }^i + {\textstyle\frac{1}{2}}\,\varPhi^{\lambda}_i
\,\varPhi^{\lambda^\prime}_j\,M_{\lambda\lambda^\prime}\, T^{ij} \,,
\nonumber
\\
L^{-1} T_{\lambda }^i L &=& \left( \delta_\lambda^{\lambda^\prime}- C
M_\lambda{}^{\lambda^\prime} \right) \,T_{\lambda^\prime }^i +
T^{ij}\,\varPhi_{j\lambda} \,,
\nonumber
\\
L^{-1} T^{ij} L &=& T^{ij} \,,
\end{eqnarray}
and, thus, read the non--vanishing components of the Killing vectors
\begin{eqnarray}
k_0 &=& \partial_0 + \varPhi^{\lambda }_i \,M_\lambda{}^{\lambda^\prime}\,
\partial_{\lambda^\prime }^i \,,
\nonumber
\\
k_{\lambda }^i &=& \partial_{\lambda }^i+{\textstyle\frac{1}{2}}\,
\varPhi_{j\lambda}\,\partial^{ij} \,, 
\nonumber
\\
k^{ij} &=& \partial^{ij} \,,
\end{eqnarray}
where
\begin{equation}
\partial_0 = {\partial \over \partial C}\,, \qquad \partial^{ij} = {\partial
\over \partial C_{ij}} \quad {\rm and} \qquad \partial^i_\lambda = {\partial 
\over \partial \varPhi^\lambda_i} \,.
\end{equation}
Therefore, under the infinitesimal diffeomorphism $\xi^0 k_0+\xi^{\lambda }_i 
k_{\lambda }^i + \xi_{ij} k^{ij}$ the fields transform as follows:
\begin{eqnarray}
\delta C &=& \xi^0 \,,
\nonumber
\\
\delta\varPhi^{\lambda }_i &=& \xi^{\lambda}_i+\xi^0\,
\varPhi^{\lambda^\prime }\,M_{\lambda^\prime}{}^{\lambda} \,,
\nonumber
\\
\delta C_{ij} &=& \xi_{ij} + {\textstyle\frac{1}{2}}\,
\xi^{\lambda}_{[i}\,\varPhi_{j]\lambda} \,.
\end{eqnarray}

\noindent{\it Scalar kinetic terms. \ \ }
Since all the quantities of our gauging are covariant with respect
to $SO(2,2)\times GL(4,\mathbb{R})$ it is useful to define the
(full) coset representative in the following way
\begin{equation}
L = \exp{\left(C_{ij}\,T^{ij}\right)}\,
\exp{\left(\varPhi^{\lambda}_i\,T_{\lambda }^i\right)}\exp{\left(c\,
T\right)}\,\mathbb{E} \,,
\end{equation} 
where $\mathbb{E}$ is the coset representative of the submanifold
\begin{equation}
\mathbb{E} \in  O(1,1)_0\times \frac{SO(2,2)}{SO(2)\times
SO(2)}\times \frac{GL(4,\mathbb{R})}{SO(4)} \,. \label{subman}
\end{equation}
The scalar kinetic terms are computed by evaluating the components
of the vielbein ${\Scr P}\,=\, L^{-1}dL_{|G/H}$:
\begin{equation}
L^{-1}dL_{|G/H} = {\Scr P}_{\hat{\imath}\hat{\jmath}}\,
\hat{T}^{\hat{\imath}\hat{\jmath}}+
{\Scr P}_{\hat{\imath}}^{\hat{\lambda}}\,
\hat{T}^{\hat{\imath}}_{\hat{\lambda}}+ {\Scr P}\,
\hat{T}+ {\Scr P}_E \,,
\end{equation}
where the restriction to $G/H$ amounts to select the
non--compact isometries of the scalar manifold, ${\Scr P}_E$
is the algebra--valued vielbein of the submanifold (\ref{subman}).
Finally, the hatted generators denote the non--compact
component of the corresponding solvable generator. The kinetic
Lagrangian for the scalar fields is then 
\begin{equation}
{\Scr L}_{\rm scal} =
{\textstyle\frac{1}{2}}\, {\Scr P}_{\mu}\,
{\Scr P}^\mu+{\textstyle\frac{1}{2}}\,\sum_{\hat{\imath}\hat{\lambda}}\,
{\Scr P}_{\hat{\imath}\,\mu}^{\hat{\lambda}}\,
{\Scr P}_{\hat{\imath}}^{\hat{\lambda}\,\mu}\,+
{\textstyle\frac{1}{4}}\,\sum_{\hat{\imath}\hat{\jmath}}\,
{\Scr P}_{\hat{\imath}\hat{\jmath}\,\mu}\, 
{\Scr P}_{\hat{\imath}\hat{\jmath}}{}^\mu+
{\rm Tr}( {\Scr P}_E^2) \,,
\end{equation}
where
\begin{eqnarray}
{\Scr P}_{\mu} &=& \partial_\mu c \,,
\nonumber \\
{\Scr P}_{\hat{\imath}\mu}^{\hat{\lambda}} &=&
(\partial_\mu \varPhi^{\lambda}_i)\,E^i{}_{\hat{\imath}}\,
E_\lambda{}^{\hat{\lambda}}\,,
\nonumber\\
{\Scr P}_{\hat{\imath}\hat{\jmath}\mu}&=&\left[\partial_\mu
C_{ij}+{\textstyle\frac{1}{4}} (\partial_\mu
\varPhi_i^\lambda\,\varPhi_{j\lambda}-\partial_\mu
\varPhi_j^\lambda\,\varPhi_{i\lambda})\right]\,
E^i{}_{\hat{\imath}}\,E^j{}_{\hat{\jmath}} \,.
\end{eqnarray}

\noindent{\it Vector fields.\ \ } The twelve vector potentials are
$B_{a\mu},\,C_{a\mu},\,G^i_\mu,\,C_{ijk\mu}$. As before, we shall collectively
denote by $A^\lambda_\mu $ the pair $\{B_{a\mu},\,C_{a\mu}\}$, and by
$F^\lambda=dA^\lambda $ the corresponding field strengths. To avoid confusion,
we shall then adopt the following notation for the remaining field strengths: 
${\Scr F}^i=dG^i$ and $F^i=\epsilon^{ijkl}\,dC_{jkl}$. Moreover, 
$\tilde{F}_\lambda,\,\tilde{{\Scr F}}_i $ and $\tilde{F}_i$ will denote 
the ``dual'' field strengths, obtained by varying the Lagrangian
with respect to the electric ones, not to be confused with the 
four--dimensional Hodge duals ${}^*{F}^\lambda,\,{}^*{{\Scr F}}^i $ and
${}^*{F}^i$. Following \cite{Gaillard:1981rj}, we can then collect the 
field strengths and their duals in a symplectic vector
\begin{equation}
\{ F^\lambda,\,{\Scr F}^i,\, F^i,\,\tilde{F}_\lambda,\, 
\tilde{{\Scr F}}_i,\,\tilde{F}_i \}\label{symba} \,.
\end{equation}
In table 6, we list the field strengths and their duals as 
they appear in the symplectic section, together with their $O(1,1)^3$ gradings 
and the corresponding weights of the ${\bf 56}$ of $E_{7(7)}$.

\begin{table}[ht]
\vskip 0.5 cm
\caption{Field strengths, $O(1,1)^3$ gradings, and corresponding weights.}
\vskip 0.5 cm
\begin{center}
\begin{tabular}{|c|c|c|}
\hline
Sp--section  & $ O(1,1)^3$--grading & weight 
\\\hline
$F_{1a}$ & $(-1,0,-{1\over 2})$ & $\epsilon_a- {1\over \sqrt{2}} \epsilon_7$
\\\hline
$F_{2a}$ & $(-1,0,{1 \over 2})$ & $w + \epsilon_a$
\\\hline
${\Scr F}^i$  & $(-1,-1,-{1 \over 2})$ & $-\epsilon_i- {1\over \sqrt{2}}
\epsilon_7$
\\\hline
$F^i$  & $(1,-1,-{1 \over 2})$ & $w+ \epsilon_j+ \epsilon_k+ \epsilon_l$
\\\hline
$\tilde{F}^{1a}$ & $(1,0,{1 \over 2})$ & $-\epsilon_a+ {1\over\sqrt{2}}
\epsilon_7$ 
\\\hline
$\tilde{F}^{2a}$ & $(1,0,-{1 \over 2})$ & $ -w - \epsilon_a$ 
\\\hline
$\tilde{{\Scr F}}_{i}$ & $(1,1,{1 \over 2})$ & $\epsilon_i+
{1\over \sqrt{2}} \epsilon_7$ 
\\\hline
$\tilde{F}_{ i}$ & $(-1,1,{1 \over 2})$ & $-w- \epsilon_j- \epsilon_k- 
\epsilon_l$ 
\\\hline
\end{tabular}
\vskip 0.5 cm
\end{center}
\end{table}

Under a generic nilpotent transformation 
\begin{equation}
\xi T+\xi_0\, T_0+\xi^\lambda_i T_\lambda^i+\xi_{ij} T^{ij}\,,
\end{equation} 
the field strengths transform as
\begin{eqnarray}
\delta F^\lambda & = & -\xi^\lambda_i\, {\Scr F}^i + \xi_0 \, 
F^{\lambda^\prime}\, M_{\lambda^\prime}{}^\lambda \,,
\nonumber \\
\delta {\Scr F}^i &=& 0 \,,
\nonumber
\\
\delta F^i & = & \xi\,{\Scr F}^i \,,
\nonumber 
\\
\delta\tilde{F}_\lambda &=& \xi\,\eta_{\lambda\lambda^\prime}\,
F^{\lambda^\prime}- \xi_0\, M_\lambda{}^{\lambda^\prime}\,
\tilde{F}_{\lambda^\prime}-\eta_{\lambda\lambda^\prime}\,
\xi_i^{\lambda^\prime}\,F^i \,,
\nonumber
\\
\delta\tilde{{\Scr F}}_i & = & -\xi\,\tilde{F}_i+\xi^\lambda_i\,
\tilde{F}_{\lambda}- 2\,\xi_{ij}\,F^j\,,
\nonumber 
\\
\delta\tilde{F}_i & = & -\xi_i^{\lambda^\prime}\,
F^{\lambda}\,\eta_{\lambda\lambda^\prime}+2\,\xi_{ij}\,{\Scr F}^j \,.
\end{eqnarray}

We then deduce that the electric subalgebra is
\begin{equation}
g_e = {o}(1,1)_{(0,0)}+ so (2,2)_{(0,0)} + gl (4,\mathbb{R})_{(0,0)} + {\bf
(1,1)}_{(2,0)}+{\bf (4,4)}_{(0,1)}+{\bf (1,6)}_{(0,2)} \,,
\nonumber
\end{equation}
where ${o}(1,1)_{(0,0)}$ is the generator of $O(1,1)_0$, and the grading 
refers to $O(1,1)_0\times O(1,1)_1$. The group $O(1,1)_2$ is now included 
inside $SO(2,2)$ and, in what follows, we shall not consider its grading
any longer. Furthermore, we identify $T$ as the generator in 
${\bf (1,1)}_{(2,0)}$, $T^i_\lambda$ and $T^{ij}$ are associated to 
${\bf (4,4)}_{(0,1)}$ and ${\bf (6,1)}_{(0,2)}$, respectively.
The interested reader may find in appendix the explicit symplectic 
realisation of the generators of $N_7$, as well as the computation of the 
vector kinetic matrix.

\subsection{$T_2 \times T_4$ IIB orientifold with D5 branes.
The ungauged version}

\noindent{\it Solvable algebra of global symmetries.\ \ } 
In this second model the relevant axions are $B_{ia},\,C_{ab},\,C_{iabc}\equiv
C_i^d,\,C_{\mu\nu}\equiv c$ and $C_{ij}=\epsilon_{ij}\,c^\prime$, and can be
associated to the following choice of simple roots
\begin{eqnarray}
\beta_1 &=& \epsilon_1-\epsilon_2\,,
\nonumber \\
\beta_2 &=& \epsilon_2+\epsilon_3\,,
\nonumber\\
\beta_3 &=& -\epsilon_3+\epsilon_4\,,
\nonumber \\
\beta_4 &=& -\epsilon_4+\epsilon_5\,,
\nonumber \\
\beta_5 &=& -\epsilon_5+\epsilon_6\,,
\nonumber \\
\beta_6 &=& a+\epsilon_3+\epsilon_4\,,
\nonumber
\end{eqnarray}
for the subalgebra $so(6,6) \subset e_{7(7)}$. The Dynkin diagram of the 
external $SL(2,\mathbb{R})$ consists, instead, of the single root
\begin{equation}
\beta = a + \epsilon_1 + \epsilon_2 \,,
\end{equation}
whose corresponding axion is $C_{ij}$, according to eq. (\ref{eq1}). 

The triple grading, this time, refers to the $O(1,1)^3$ group generated by 
the three Cartan $H_{\beta},\, H_{\lambda^2},\,H_{\lambda^6}$ and parametrised 
by the  moduli
$\beta\cdot h,\,h_2,\,h_6$:
\begin{eqnarray}
O(1,1)_0 &\rightarrow &  e^{\beta\cdot h}\,=\,V_2\, e^{-\frac{\phi}{2}} \,,
\nonumber\\
O(1,1)_1&\rightarrow& e^{h_2}\,=\,(V_2)^{\frac{1}{2}}\,(V_4)^{\frac{1}{4}}
\,e^{\frac{\phi}{2}} \,,
\nonumber\\
O(1,1)_2&\rightarrow &e^{h_6}\,=\,(V_4)^{\frac{1}{2}}\,e^{-\frac{\phi}{2}} \,,
\end{eqnarray}
where, as usual,  $O(1,1)_0$ is in the external $SL(2,\mathbb{R})$, while 
$O(1,1)_1$ and 
$O(1,1)_2$ are contained in $GL(2,\mathbb{R})\times GL(4,\mathbb{R})$.

In table 7 we list the axionic fields of this model, together with the 
corresponding generator of $Solv (so(6,6))$, for each of which the 
$O(1,1)^3$ grading is specified, as well as their $SL(2,\mathbb{R}) \times
SL(4,\mathbb{R})$ representations

\begin{table}[ht]
\vskip 0.5 cm 
\caption{Axionic fields for the $T_2\times T_4$ IIB orientifold, generators
of $Solv(so(6,6))$, $O(1,1)^3$ gradings, and $SL(2, \mathbb{R}) \times 
SL(4,\mathbb{R})$ representations.}
\vskip 0.5 cm
\begin{center}
{\small \begin{tabular}{|c|c|c|c|c|}\hline
$GL(2)\times GL(6)$--rep. & generator & root & field & dim. 
\\\hline
--- & $T_{(0,0,0)}$ & $\{\epsilon_i-\epsilon_j,\,\epsilon_a-\epsilon_b\}$ 
{\scriptsize ($i<j,\,a>b$)} & $\{g_{ij},\,g_{ab}\}$ & 7 
\\\hline
${\bf (1,6)}_{(0,0,1)}$ & $T^{ab}$ &$\alpha_7+\epsilon_a+\epsilon_b$ & 
$C_{ab}$ & 6  
\\\hline
${\bf (2,4)}_{(0,1,0)}$  & $T^{ia}$ & $\epsilon_i+\epsilon_a $& $B_{ia}$ & 8 
\\\hline
${\bf (2,\overline{4})}_{(0,1,1)}$ & $T^{i}_d$ &$\alpha_7 +\epsilon_i+
\epsilon_a+\epsilon_b+\epsilon_c$ &   $C_{i}^d$ & 8 
\\\hline
${\bf (1,1)}_{(0,2,1)}$ & $T$ & $\alpha_7+\epsilon_i+\epsilon_j+\epsilon_a+
\epsilon_b+ \epsilon_c+\epsilon_d$ &$C_{\mu\nu}\,=\, c$ & 1 
\\
\hline\hline
${\bf (1,1)}_{(2,0,0)}$ & $T^\prime$ & $\beta$ &$C_{ij}\,=\, c^\prime$ & 1 
\\\hline
\end{tabular}}
\vskip 0.5 cm
\end{center}
\end{table}

Also in this case, the generators $T$, $T^{ia}$, $T^i_a$ and $T^{ab}$ close a
23--dimensional solvable subalgebra of $SO(6,6)$
\begin{equation}
N_5 = c \,T+B_{ia}\,T^{ia}+C_i^a\,T^i_a+C_{ab}\,T^{ab} \,,
\end{equation}
whose algebraic structure is encoded in the non--vanishing commutators
\begin{eqnarray}
\label{comm1} \left[T^{ia},\,T^{bc}\right] & = & \epsilon^{abcd}\,T^i_d \,,
\\
\label{com2} \left[T^{ia},\,T^j_d\right] &=& \epsilon^{ij}\,\delta^a_d\,T \,.
\end{eqnarray}
The corresponding coset representative reads
\begin{equation}
L = e^{c\,T}\,e^{B_{ia}\,T^{ia}}\,e^{C_i^a\,T^i_a}\,e^{C_{ab}\,T^{ab}} \,,
\end{equation}
while its left--invariant one--form is
\begin{equation}
L^{-1}dL =  T dc+T^{ab} dC_{ab}+T^i_d\,dC^d_i+
(T^{ia}+\epsilon^{ij}\,C_j^a\,T+\epsilon^{abcd}\,T^i_d\,C_{bc})\,dB_{ia} \,.
\end{equation}
The transformation properties of the axionic scalars can be deduced from 
\begin{eqnarray}
L^{-1}TL &=& T \,,
\nonumber\\
L^{-1}T^i_a L &=& T^i_a +\epsilon^{ij}\, B_{ja}\, T \,,
\nonumber\\
L^{-1}T^{ia} L &=& T^{ia}+\epsilon^{ij}\,C_j^a\,T+\epsilon^{abcd}\,T^i_d\,
C_{bc} \,,
\nonumber\\
L^{-1}T^{ab}L &=& T^{ab}+\epsilon^{abcd}\,B_{id}\,T^i_c \,,
\end{eqnarray}
which identify the Killing vectors
\begin{eqnarray}
k &=& \partial \,,
\nonumber \\
k^i_a &=& \partial^i_a + \epsilon^{ij} B_{ja} \partial \,,
\nonumber \\
k^{ia} &=& \partial^{ia} \,,
\nonumber \\
k^{ab} &=& \partial^{ab} +\epsilon^{abcd} B_{id} \partial^i_c \,,
\end{eqnarray}
where 
\begin{equation}
\partial = {\partial \over \partial c}\,, \qquad 
\partial^i_a = {\partial \over \partial C_i^a}\,, \qquad
\partial^{ia} =  {\partial \over \partial B_{ia}}\,, \qquad
\partial^{ab} =  {\partial \over \partial C_{ab}}\,.
\end{equation}
Hence, under the infinitesimal diffeomorphism $\xi \,T+\xi_{ia}\,T^{ia}+
\xi_i^a\,T^i_a+\xi_{ab}\,T^{ab}$, one has
\begin{eqnarray}
\delta c &=& \epsilon^{ij}\,\xi_{i}^a\,B_{ja}+\xi\,,
\nonumber\\
\delta C_i^a &=& \epsilon^{abcd}\,\xi_{bc}\,B_{id}+\xi_i^a \,,
\nonumber\\
\delta B_{ia}&=&\xi_{ia} \,,
\nonumber\\
\delta C_{ab}&=&\xi_{ab} \,.
\end{eqnarray}
For later convenience we shall define the generator $T_{ab}=-\frac{1}{4}\,
\epsilon_{abcd}\, T^{cd}$, and the corresponding parameter $\xi_{ab} =
-\frac{1}{4}\, \epsilon_{abcd}\, \xi^{cd}$, in terms of which the relation 
(\ref{comm1}) reads
\begin{equation}
\label{com1p} \left[T_{ab},\, T^{ic}\right] = \delta^c_{[a}\, T^i_{b]} \,.
\end{equation}

\noindent{\it Vector fields.\ \ } The vector fields of this model 
are $G^i_\mu,\, C_{i\mu}\, B_{a\mu},\, C^a_\mu$, and we name the corresponding 
field strengths and their duals by
\begin{equation}
{\Scr F}^i_{\mu\nu},\,F_{i\mu\nu},\,{\Scr H}_{a\mu\nu},\,F^a_{\mu\nu},\,
\tilde{{\Scr F}}_{i\mu\nu},\,\tilde{F}^i_{\mu\nu},\,
\tilde{{\Scr H}}^a_{\mu\nu},\,\tilde{F}_{a\mu\nu} \,.
\label{symba2}
\end{equation}
In the table 8 we list the field strengths and their duals as they appear in
the symplectic section, together with their $O(1,1)^3$ gradings, and the
corresponding $E_{7(7)}$ weights.

\begin{table}[ht]
\vskip 0.5 cm 
\caption{Field strengths, $O(1,1)^3$ gradings, and corresponding weights.}
\vskip 0.5 cm
\begin{center}
\begin{tabular}{|c|c|c|}\hline
Sp--section & $O(1,1)^3$--grading & weight 
\\\hline
${\Scr F}^i_{\mu\nu}$ & $(-1,-1,-{1\over 2})$& $-\epsilon_i-{1\over \sqrt{2}}
\epsilon_7$ 
\\\hline
$F_{i\mu\nu}$ &$(1,-1,-{1\over 2})$& $w+\epsilon_i$   
\\\hline
${\Scr H}_{a\mu\nu}$  & $(-1,0,-{1\over 2})$& $\epsilon_a- {1\over\sqrt{2}}
\epsilon_7$  
\\\hline
$F^a_{\mu\nu}$  & $(-1,0,{1 \over 2})$ & $w+\epsilon_b+\epsilon_c+\epsilon_d$
\\\hline
$\tilde{{\Scr F}}_{i\mu\nu}$ & $(1,1,{1\over 2})$& $\epsilon_i+{1\over\sqrt{2}}
\epsilon_7$ 
\\\hline
$\tilde{F}^i_{\mu\nu}$ &$(-1,1,{1\over 2})$& $w+\epsilon_j+\epsilon_a+
\epsilon_b+ \epsilon_c+\epsilon_d$
\\\hline
$\tilde{{\Scr H}}^a_{\mu\nu}$  & $(1,0,{1\over 2})$ & $-\epsilon_a+
{1\over \sqrt{2}} \epsilon_7$
\\\hline
$\tilde{F}_{a\mu\nu}$  & $(1,0,-{1\over 2})$& $w+\epsilon_i+\epsilon_j+
\epsilon_a$ 
\\\hline
\end{tabular}
\vskip 0.5 cm
\end{center}
\end{table}

The transformation laws under a generic nilpotent transformation
$\xi^\prime \, T^\prime+\xi \, T+\xi^{ab}\, T_{ab}+\xi_{ia}\, T^{ia}+
\xi^a_i\, T^i_a$ can be deduced from the grading and weight structures. 
One finds
\begin{eqnarray}
\delta{\Scr F}^i &=& 0\,,
\nonumber \\
\delta F_i &=& \xi^\prime\, \epsilon_{ij}\,{\Scr F}^j \,,
\nonumber \\
\delta {\Scr H}_a &=& \xi_{ia} \,{\Scr F}^i\,,
\nonumber \\
\delta F^a &=& \xi^{ab} \,{\Scr H}_b-\xi^a_i\,{\Scr F}^i \,,
\nonumber \\
\delta\tilde{{\Scr F}}_i &=& \xi^\prime\,\epsilon_{ij}\,\tilde{F}^j+\xi^a_i\,
\tilde{F}_a-\xi_{ia}\,\tilde{{\Scr H}}^a+\xi\,F_i\,,
\nonumber \\
\delta \tilde{F}^i &=& \epsilon^{ij}\,\xi_{ja}\, F^a-\epsilon^{ij}\,
\xi_{j}^a\, {\Scr H}_a+\xi\, {\Scr F}^i
\nonumber \\
\delta \tilde{{\Scr H}}^a &=& \xi^\prime \,F^a+\xi^{ab}\,\tilde{F}_b+\xi^a_i\,
\epsilon^{ij}\,F_j\,,
\nonumber \\
\delta\tilde{F}_a &=& \xi^\prime \,{\Scr H}_a-\xi_{ai}\,\epsilon^{ij}\,F_j \,.
\label{isom}
\end{eqnarray} 
The explicit symplectic representation
of the $N_5$ generators together with the computation of the
vector kinetic matrix ${\Scr N}$ may be found in appendix.

\subsection{$T_0\times T_6$ and $T_6\times T_0$ IIB orientifolds with D3 and D9
branes. The ungauged version.}

The $T_0\times T_6$ model in the presence of $D3$--branes, with and without 
fluxes was constructed in \cite{D'Auria:2003jk, D'Auria:2002th, 
D'Auria:2002tc}. The structure of the $T_6\times T_0$ model, on the other hand,
is somewhat trivial, since there is no room for fluxes to be turned on. 
For completeness, here we shall confine ourselves to the description of their 
embeddings within the ${\Scr N}=8$ theory, and to the identification of
the solvable algebras $N_3$ and $N_9$, together with their action on scalar 
and vector fields.

\noindent{\it Solvable algebra of global symmetries: the $T_0\times T_6$ model.
\ \ } The embedding of the $sl(2,\mathbb{R})+so(6,6)$ algebra inside 
$e_{7(7)}$ is defined by the following identification of the simple roots:
\begin{eqnarray}
\beta_1 &=& -\epsilon_1+\epsilon_2\,,
\nonumber \\
\beta_2 &=& -\epsilon_2+\epsilon_3 \,,
\nonumber \\
\beta_3 &=& -\epsilon_3+\epsilon_4\,,
\nonumber\\
\beta_4 &=& -\epsilon_4+\epsilon_5\,,
\nonumber \\
\beta_5 &=& -\epsilon_5+\epsilon_6\,,
\nonumber \\
\beta_6 &=& a+\epsilon_1+\epsilon_2+\epsilon_3+\epsilon_4\,,
\end{eqnarray}
for the $so(6,6)$ component, and
\begin{equation}
\beta\,=\,a\,,
\end{equation}
for the $sl(2,\mathbb{R})$ one.
The correspondence axion--root is quite simple and is summarised in table 9.

\begin{table}[ht]
\vskip 0.5 cm
\caption{Axionic fields for the $T_0 \times T_6$ IIB orientifold, generators of
$Solv(so(6,6))$, $O(1,1)^2$ gradings and $GL(6,\mathbb{R})$ representations.}
\begin{center}
\vskip 0.5 cm
\begin{tabular}{|c|c|c|c|c|}\hline
$ GL(6)$--rep. & generator & root & field & dim. 
\\\hline
--- & $T_{(0,0)}$ & $\{\epsilon_a-\epsilon_b\}$ {\scriptsize ($a>b$)} & 
$\{g_{ab}\}$ & 15 
\\\hline
$\overline{{\bf 15}}_{(0,1)}$ & $T_{ab}$ &$a+\epsilon_{c}+\epsilon_d+
\epsilon_e+\epsilon_f$ & $ C^{ab}\,\equiv\,\epsilon^{abcdef}\,C_{cdef}$ & 15  
\\\hline\hline
${\bf 1}_{(2,0)}$ & $T$ & $\beta$ &$C_{0}\,=\, c $ & 1 
\\\hline
\end{tabular}
\vskip 0.5 cm
\end{center}
\end{table}

In this case, the grading is with respect to the pair of $O(1,1)$ groups 
generated by $H_{\beta},\, H_{\lambda^6}$ and corresponding to the following 
moduli:
\begin{eqnarray}
O(1,1)_0 &\rightarrow & e^{\beta\cdot h}\,=\,e^{-\phi} \,,
\nonumber\\
O(1,1)_1 &\rightarrow & e^{\lambda^6\cdot h}\,=\,V_6 \,.
\end{eqnarray} 
The nilpotent algebra $N_3$, generated by $T_{ab}$, acts as Peccei--Quinn 
translations on the R--R scalars $C^{ab}$
\begin{equation}
\delta C^{ab} = \xi^{ab} \,.
\end{equation}

The vector fields are $C_{a\mu}$ and $B_{a\mu}$, and the symplectic section 
of the corresponding field strengths $F_{a\mu\nu}$ and ${\Scr H}_{a\mu\nu}$ 
and their magnetic duals $\tilde{F}^a_{\mu\nu}$, $\tilde{{\Scr H}}^a_{\mu\nu}$ 
is listed in table 10.

\begin{table}[ht]
\begin{center}
\vskip 0.5 cm
\caption{Field strengths, $O(1,1)^2$ gradings, and corresponding weights.}
\vskip 0.5 cm
\begin{tabular}{|c|c|c|}\hline
Sp--section & $O(1,1)^2$--grading & weight 
\\\hline
$F_{a\mu\nu}$ & $(1,-{1\over 2})$& $w+\epsilon_a$ 
\\\hline
${\Scr H}_{a\mu\nu}$  & $(-1,-{1\over 2})$ & $\epsilon_a-{1\over \sqrt{2}}
\epsilon_7$
\\\hline
$\tilde{F}^a_{\mu\nu}$ & $(-1,{1\over 2})$& $-w-\epsilon_a$ 
\\\hline
$\tilde{{\Scr H}}^a_{\mu\nu}$  & $(1,{1\over 2})$& $-\epsilon_a+{1\over
\sqrt{2}} \epsilon_7$ 
\\\hline
\end{tabular}
\vskip 0.5 cm
\end{center}
\end{table}

The duality action of an infinitesimal transformation $\xi^{ab}\,T_{ab}+\xi\, 
T$ is then
\begin{eqnarray}
\delta F_{a} &=& \xi\,{\Scr H}_a\,,
\nonumber \\
\delta{\Scr H}_a &=& 0\,,
\nonumber\\
\delta\tilde{F}^a &=& \xi^{ab}\,{\Scr H}_{b}\,,
\nonumber \\
\delta\tilde{{\Scr H}}^a &=& -\xi^{ab}\,F_{b}-\xi\, \tilde{F}^a \,.
\end{eqnarray}

\noindent{\it Solvable algebra of global symmetries: the $T_6\times T_0$ 
model.\ \ } The embedding of the $sl(2,\mathbb{R})+so(6,6)$ algebra inside 
$e_{7(7)}$ is defined by the following identification of the simple roots
\begin{eqnarray}
\beta_1 &=&\epsilon_1-\epsilon_2\,,
\nonumber \\
\beta_2 &=&\epsilon_2-\epsilon_3\,,
\nonumber \\
\beta_3 &=& \epsilon_3-\epsilon_4
\nonumber \\
\beta_4 &=& \epsilon_4-\epsilon_5\,,
\nonumber \\
\beta_5 &=& \epsilon_5-\epsilon_6\,,
\nonumber \\
\beta_6 &=& a+\epsilon_5+\epsilon_6\,,
\end{eqnarray}
for the $so(6,6)$ component, and
\begin{equation}
\beta = a+\sum_{n=1}^6\,\epsilon_n \,,
\end{equation}
for the $sl(2,\mathbb{R})$ one.
The correspondence axion--root is quite simple, and is summarised in table 11.

\begin{table}[ht]
\vskip 0.5 cm
\begin{center}
\caption{Axionic fields for the $T_6 \times T_0$ IIB orientifold, generators of
$Solv(so(6,6))$, $O(1,1)^2$ gradings, and $GL(6,\mathbb{R})$ representations.}
\vskip 0.5 cm
\begin{tabular}{|c|c|c|c|c|}\hline
$ GL(6)$--rep. & generator & root & field & dim. 
\\\hline
--- & $T_{(0,0)}$ & $\{\epsilon_i-\epsilon_j\}$ {\scriptsize ($i<j$)} & 
$\{g_{ij}\}$ & 15 
\\\hline
${\bf 15}_{(0,1)}$ & $T^{ij}$ &$a+\epsilon_{i}+\epsilon_j$ &  $ C_{ij}$ & 15  
\\\hline\hline
${\bf 1}_{(2,0)}$ & $T$ & $\beta$ &$C_{\mu\nu}\,=\, c $ & 1 
\\\hline
\end{tabular}
\end{center}
\vskip 0.5 cm
\end{table}

In this case, the grading is with respect to a pair of $O(1,1)$ groups 
generated by $H_{\beta},\,H_{\lambda^6}$ and corresponding to the following 
moduli:
\begin{eqnarray}
O(1,1)_0 &\rightarrow & e^{\beta\cdot h}\,=\,V_6\,e^{\frac{\phi}{2}} \,,
\nonumber\\
O(1,1)_1 &\rightarrow & e^{\lambda^6\cdot h}\,=\,(V_6)^{\frac{1}{2}}\, 
e^{-\frac{3}{4}\,\phi} \,.
\end{eqnarray} 
The nilpotent algebra $N_9$, generated by $T^{ij}$, acts as Peccei--Quinn
translations on the R--R scalars $C_{ij}$,
\begin{equation}
\delta C_{ij} = \xi_{ij} \,.
\end{equation}

The vector fields are $C_{i\mu}$ and $G^i_{\mu}$, and the symplectic sections
of the corresponding field strengths $F_{i\mu\nu}$ and ${\Scr F}^i_{\mu\nu}$ 
and their magnetic duals $\tilde{F}^i_{\mu\nu}$, $\tilde{{\Scr F}}_{i\mu\nu}$ 
are listed in table 12.

\begin{table}[ht]
\begin{center}
\vskip 0.5 cm
\caption{Field strengths, $O(1,1)^2$ gradings, and corresponding weights.}
\vskip 0.5 cm
\begin{tabular}{|c|c|c|}\hline
Sp--section & $O(1,1)^2$--grading & weight 
\\\hline
$F_{i\mu\nu}$ & $(-1,{1\over 2})$& $w+\epsilon_i$ 
\\\hline
${\Scr F}^i_{\mu\nu}$  & $(-1,-{1\over 2})$ & $-\epsilon_i-
{1\over\sqrt{2}}\epsilon_7$
\\\hline
$\tilde{F}^i_{\mu\nu}$ & $(1,-{1\over 2})$& $-w-\epsilon_i$ 
\\\hline
$\tilde{{\Scr F}}_{i\mu\nu}$  & $(1,{1\over 2})$& $\epsilon_i+
{1\over\sqrt{2}}\epsilon_7$ 
\\\hline
\end{tabular}
\vskip 0.5 cm
\end{center}
\end{table}

The duality action of an infinitesimal transformation $\xi_{ij}\,T^{ij}+\xi\, 
T$ is then
\begin{eqnarray}
\delta F_{i} &=& \xi_{ij}\,{\Scr F}^j\,,
\nonumber \\
\delta{\Scr F}^i &=& 0 \,,
\nonumber\\
\delta\tilde{F}^i &=& \xi\,{\Scr F}^i\,,
\nonumber \\
\delta\tilde{{\Scr F}}_i &=& \xi_{ij}\,\tilde{F}^j+\xi\,F_{i} \,. 
\end{eqnarray}
As a result, the electric group contains the whole $SO(6,6)$, as for the 
heterotic string on $T_6$. In other words, there are no Peccei--Quinn 
isometries in $SO(6,6)$ which could be gauged. This feature is consistent with 
the fact that this model does not allow fluxes, and usually fluxes translate
into local Peccei--Quinn invariances in the low--energy supergravity 
description.  

\subsection{$T_1\times T_5$ IIA orientifold with $D4$--branes.} 

\noindent{\it Solvable algebra of global symmetries.\ \ } The embedding of the
$sl(2,\mathbb{R}) + so(6,6)$ algebra inside $e_{7(7)}$ is defined by the 
following identifications of simple roots:
\begin{eqnarray}
\beta_1 &=& \epsilon_1+\epsilon_2\,,
\nonumber \\
\beta_2 &=& -\epsilon_2+\epsilon_3\,,
\nonumber\\
\beta_3 &=& -\epsilon_3+\epsilon_4\,,
\nonumber \\
\beta_4 &=& -\epsilon_4+\epsilon_5\,,
\nonumber\\
\beta_5 &=& -\epsilon_5+\epsilon_6\,,
\nonumber \\
\beta_6 &=& a+\epsilon_2+\epsilon_3+\epsilon_4\,,
\nonumber\\
\end{eqnarray}
for the $so(6,6)$ factor, and
\begin{equation}
\beta = a+\epsilon_1 \,,
\end{equation}
for the $sl(2,\mathbb{R})$ one. The correspondence axion--root is quite simple,
and is summarised in table 13.

\begin{table}[ht]
\vskip 0.5 cm
\caption{Axionic fields for the $T_1\times T_5$ IIA orientifold, generators
of $Solv(so(6,6))$, $O(1,1)^3$ gradings, and $GL(5,\mathbb{R})$ 
representations.}
\begin{center}
\vskip 0.5 cm
\begin{tabular}{|c|c|c|c|c|}\hline
$ GL(5)$--rep. & generator & root & field & dim. 
\\\hline
--- & $T_{(0,0,0)}$ & $\{\epsilon_a-\epsilon_b\}$ {\scriptsize ($a>b$)} & 
$\{g_{ab}\}$ & 10 
\\\hline
$\overline{{\bf 10}}_{(0,0,1)}$ & $T_{ab}$ &$a+\epsilon_c+\epsilon_d+
\epsilon_e$ & $C_{cde}\,\equiv\, C^{ab}$ & 10  
\\\hline
${\bf 5}_{(0,1,0)}$  & $T^{a}$ & $\epsilon_1+\epsilon_a $& $B_{1a}\,\equiv\, 
B_a$ & 5 
\\\hline
$\overline{{\bf 5}}_{(0,1,1)}$ & $T_e$ &$ a +\epsilon_1+\epsilon_a+\epsilon_b+
\epsilon_c+\epsilon_d $ & $C_{\mu\nu a}\,\equiv\, C^e$ & 5 
\\\hline\hline
${\bf 1}_{(2,0,0)}$ & $T$ & $\beta$ &$C_{1}\,=\, c $ & 1 
\\\hline
\end{tabular}
\vskip 0.5 cm
\end{center}
\end{table}

In this case the grading is with respect to the $O(1,1)^3$ group generated by 
 $H_{\beta},\,H_{\lambda^1},\,H_{\lambda^6}$  and parametrised by the  moduli
$\beta\cdot h,\,h_1,\,h_6$: 
\begin{eqnarray}
O(1,1)_0 &\rightarrow &  e^{\beta\cdot h}\,=
\,V_1\, e^{-\frac{3}{4}\,\phi}\,,
\nonumber\\
O(1,1)_1&\rightarrow& e^{h_1}\,=
\,V_1\,(V_5)^{\frac{1}{5}}\, e^{\frac{\phi}{2}} \,,
\nonumber\\
O(1,1)_2&\rightarrow &e^{h_6}\,=\,(V_5)^{\frac{3}{2}}\, 
e^{-\frac{\phi}{4}} \,.
\end{eqnarray}

The generators $T^a$, $T_a$ and $T_{ab}$ close a twenty--dimensional nilpotent 
subalgebra $N_4$ of $Solv(so(6,6))$:
\begin{equation}
N_4 = B_{a}\,T^{a}+C^a\,T_a+C^{ab}\,T_{ab} \,,
\end{equation}
whose algebraic structure is encoded in the non--vanishing commutator
\begin{equation}
\label{com1} \left[T_{ab},\,T^{c}\right] = T_{[a}\delta^c_{b]} \,.
\end{equation}
The corresponding coset representative reads
\begin{equation}
L =  e^{C^a\,T_a}\,e^{B_{a}\,T^{a}}\,e^{C^{ab}\,T_{ab}}\,e^{c\,T}\,
\mathbb{E} \,,\label{coset1t5}
\end{equation}
where the $\mathbb{E}$ factor parametrises the submanifold:
\begin{equation}
O(1,1)_0\times O(1,1)_1\times O(1,1)_2\times \frac{SL(5,\mathbb{R})}{SO(5)} \,.
\end{equation}
A generic element $\xi_{a}\,T^{a}+\xi^a\,T_a+\xi^{ab}\,T_{ab}$ of $N_4$ then
induces the following transformations on the axionic scalars
\begin{eqnarray}
\delta C^a &=&\xi^a+\xi^{ab}\, B_b\,,
\nonumber\\
\delta B_a &=& \xi_a \,,
\nonumber\\
\delta C^{ab} &=&\xi^{ab} \,.
\end{eqnarray}

\noindent{\it Vector fields.\ \ } The vector fields of this model are
$C_\mu,\,G^1_\mu,\,C_{1a\mu},\,B_{a\mu} $, and we name the corresponding 
field strengths $F_{\mu\nu}$, ${\Scr F}^1_{\mu\nu}$, $F_{1a\mu\nu}$, 
${\Scr H}_{a\mu\nu}$. The symplectic section of the field strengths and their
duals is
\begin{equation}
\{F_{\mu\nu},
\,{\Scr F}^1_{\mu\nu},\,F_{1a\mu\nu},\,{\Scr H}_{a\mu\nu},\,\tilde{F}_{\mu\nu},
\,\tilde{{\Scr F}}_{1\mu\nu},\,\tilde{F}^{1a}_{\mu\nu},\,
\tilde{{\Scr H}}^a_{\mu\nu}\} \,,
\end{equation}
and in table 14 we give their $O(1,1)^3$ gradings and the corresponding
$E_{7(7)}$ weights.

\begin{table}[ht]
\vskip 0.5 cm
\caption{Field strengths, $O(1,1)^3$ gradings, and corresponding weights.}
\begin{center}
\vskip 0.5 cm
\begin{tabular}{|c|c|c|}\hline
vector & $O(1,1)^3$--grading & weight 
\\\hline
$F_{\mu\nu}$ & $(1,-1,-{1\over 2})$& $w$ 
\\\hline
${\Scr F}^1_{\mu\nu}$ &$(-1,-1,-{1\over 2})$& $-\epsilon_1-
{1\over \sqrt{2}}\epsilon_7$   
\\\hline
$F_{1a\mu\nu}$  & $(1,0,-{1\over 2})$& $w+\epsilon_1+\epsilon_a$  
\\\hline
${\Scr H}_{a\mu\nu}$  & $(-1,0,-{1\over 2})$ & $\epsilon_a-
{1\over\sqrt{2}}\epsilon_7$
\\\hline
$\tilde{F}_{\mu\nu}$ & $(-1,1,{1\over 2})$& $-w$ 
\\\hline
$\tilde{{\Scr F}}_{1\mu\nu}$ &$(1,1,{1\over 2})$& $\epsilon_1+
{1\over\sqrt{2}}\epsilon_7$
\\\hline
$\tilde{F}^{1a}_{\mu\nu}$  & $(-1,0,{1\over 2})$ & $-w-\epsilon_1-\epsilon_a$  
\\\hline
$\tilde{{\Scr H}}^a_{\mu\nu}$  & $(1,0,{1\over 2})$& $-\epsilon_a+
{1\over\sqrt{2}} \epsilon_7$
\\\hline
\end{tabular}
\vskip 0.5 cm
\end{center}
\end{table}

The action of infinitesimal duality transformation $\xi_{a}\,T^{a}+\xi^a\,
T_a+\xi^{ab}\,T_{ab}+\xi\, T$ on the symplectic section is
\begin{eqnarray}
\delta F &=& \xi\, {\Scr F}^1\,,
\nonumber \\
\delta {\Scr F}^1 &=&  0\,,
\nonumber \\
\delta F_{1a} &=& \xi_a\, F+\xi\, {\Scr H}_a\,,
\nonumber \\
\delta{\Scr H}_a &=& \xi_a\,{\Scr F}^1 \,,
\nonumber\\
\delta \tilde{F} &=& -\xi_a\, \tilde{F}^{1a}+\xi^a\, {\Scr H}_a\,,
\nonumber \\
\delta \tilde{{\Scr F}}_1 &=& -\xi_a\,\tilde{{\Scr H}}^a-\xi^a\,
F_{1a}-\xi\,\tilde{F}\,,
\nonumber \\
\delta\tilde{F}^{1a} &=& -\xi^a\,{\Scr F}^1-\xi^{ab}\,{\Scr H}_b \,,
\nonumber \\
\delta\tilde{{\Scr H}}^a &=&\xi^a\,F+\xi^{ab}\,F_{1b}-\xi \tilde{F}^{1a} \,.
\end{eqnarray}
The explicit symplectic realisation of the $N_4$ generators
together with the computation of the vector kinetic matrix can be
found in appendix. 

\subsection{$T_3\times T_3$ IIA orientifold with $D6$--branes.} 

\noindent{\it Solvable algebra of global symmetries.\ \ } The embedding of the 
$sl(2,\mathbb{R})+ so(6,6)$ algebra inside $e_{7(7)}$ is defined by
the following identification of the simple roots
\begin{eqnarray}
\beta_1 &=& \epsilon_1-\epsilon_2\,,
\nonumber \\
\beta_2 &=& \epsilon_2-\epsilon_3\,,
\nonumber\\
\beta_3 &=& \epsilon_3+\epsilon_4\,,
\nonumber \\
\beta_4 &=& -\epsilon_4+\epsilon_5\,,
\nonumber\\
\beta_5 &=& -\epsilon_5+\epsilon_6\,,
\nonumber \\
\beta_6 &=& a+\epsilon_4
\end{eqnarray}
for the $so(6,6)$ factor, and
\begin{equation}
\beta = a+\epsilon_1+\epsilon_2+\epsilon_3 \,,
\end{equation}
for the $sl(2,\mathbb{R})$ one. The correspondence axion--root is quite
simple and is summarised in table 15.

\begin{table}[ht]
\vskip 0.5 cm
\caption{Axionic fields for the $T_3 \times T_3$ IIA orientifold, generators
of $Solv (so(6,6))$, $O(1,1)^3$ gradings, and $GL(3,\mathbb{R}) \times
GL(3,\mathbb{R})$ representations.}
\begin{center}
\vskip 0.5 cm
\begin{tabular}{|c|c|c|c|c|}\hline
$ GL(3)\times GL(3)$--rep. & generator & root & field & dim. 
\\\hline
--- & $T_{(0,0,0)}$ & $\{\epsilon_i-\epsilon_j,\,\epsilon_a-\epsilon_b\}$ 
{\scriptsize ($i<j,\,a>b$)} & $\{g_{ij},\,g_{ab}\}$ & 6 
\\\hline
${\bf (1,3)}_{(0,0,1)}$ & $T_{ab}$ &$a+\epsilon_c$ & $C^{ab}$ & 3  
\\\hline
${\bf (3,3)}_{(0,1,0)}$  & $T^{ia}$ & $\epsilon_i+\epsilon_a $& $B_{ia}$ & 9 
\\\hline
${\bf (3,\overline{3})}_{(0,1,1)}$ & $T^i_a$ &$ a +\epsilon_i+\epsilon_b+
\epsilon_c $ & $C_{ibc}\,\equiv\, C^a_i$ & 9
\\\hline
${\bf (\overline{3},1)}_{(0,2,1)}$  & $T^{ij}$ & $\epsilon_i+\epsilon_j +
\epsilon_a+\epsilon_b+\epsilon_c$& $C_{k\mu\nu}\,\equiv\, C_{ij}$ & 3 
\\\hline\hline
${\bf (1,1)}_{(2,0,0)}$ & $T$ & $\beta$ &$C_{ijk}\,\equiv\, c $ & 1 
\\\hline
\end{tabular}
\vskip 0.5 cm
\end{center}
\end{table}

The triple grading refers to three $O(1,1)$ groups generated by $H_{\beta},
\,H_{\lambda^3},\,H_{\lambda^6}$  and parametrised by the  moduli
$\beta\cdot h,\,h_3,\,h_6$: 
\begin{eqnarray}
O(1,1)_0  &\rightarrow &  e^{\beta\cdot h}\,=\,V_3\, e^{-\frac{\phi}{4}}\,,
\nonumber\\
O(1,1)_1 &\rightarrow & e^{h_3}\,=\,(V_3)^{\frac{1}{3}}\,
(V_3^\prime)^{\frac{1}{3}}\, 
e^{\frac{\phi}{2}}\,,
\nonumber\\
O(1,1)_2 &\rightarrow & e^{h_6}\,=\,(V_3^\prime)^{\frac{1}{3}}\, 
e^{-\frac{3}{4}\,\phi} \,.
\end{eqnarray}

The generators $T^{ia}$, $T_{ab}$, $T^i_a$ and $T^{ij}$ form now a 
24--dimensional solvable subalgebra $N_6$ of $Solv(so(6,6))$: 
\begin{equation}
N_6 = B_{ia}\,T^{ia}+C^{ab}\,T_{ab}+C^a_i\, T^i_a+C_{ij}\,T^{ij} \,,
\end{equation}
whose algebraic structure is encoded in the non--vanishing commutators
\begin{eqnarray}
\label{ncom1} \left[T_{ab},\,T^{ic}\right] &=& T^i_{[a}\delta^c_{b]} \,,
\nonumber\\
\label{ncom2} \left[T_{ia},\,T^{j}_b\right] &=& T^{ij}\delta^a_{b} \,.
\end{eqnarray}
A possible choice for the coset representative is then 
\begin{equation}
L = e^{C_{ij}\,T^{ij}}\,e^{C^{a}_i\,T_{a}^i}\,e^{B_{ia}\,T^{ia}}\,e^{C^{ab}\,
T_{ab}}\,e^{c\,T}\,\mathbb{E} \,, \label{coset3t3}
\end{equation}
with $\mathbb{E}$ parameterising the submanifold
\begin{equation}
O(1,1)_0\times\frac{GL(3,\mathbb{R})}{SO(3)}\times
\frac{GL(3,\mathbb{R})}{SO(3)} \,.\label{et3t3}
\end{equation}
Under an infinitesimal transformation $\xi_{ij}\,T^{ij}+\xi^{a}_i\,T_{a}^i+
\xi_{ia}\,T^{ia}+\xi^{ab}\,T_{ab}$ of $N_6$ the variation of the axionic
scalars is
\begin{eqnarray}
\delta C^a_i &=& \xi^{a}_i+\xi^{ab}\, B_{ib} \,,
\nonumber\\
\delta C_{ij} &=& \xi_{ij}+\xi_{a[i}\,C^a_{j]} \,,
\nonumber\\
\delta B_{ia} &=& \xi_{ia} \,,
\nonumber\\
\delta C^{ab} &=& \xi^{ab} \,.
\end{eqnarray}

\noindent{\it Vector fields.\ \ } The vector fields of this model are
$G^i_\mu$,$C^i_\mu = \epsilon^{ijk}\,C_{jk\mu}$, $B_{a\mu}$, $C^a_{\mu}= 
\epsilon^{abc}\,C_{bc\mu}$, and we name the corresponding 
field strengths ${\Scr F}^i_{\mu\nu}$, $F^i_{\mu\nu}$, ${\Scr H}_{a\mu\nu}$, 
$F^a_{\mu\nu}$. The symplectic section of the field strengths and their duals
is
\begin{equation}
\{{\Scr F}^i_{\mu\nu}, \,F^i_{\mu\nu},\,{\Scr H}_{a\mu\nu},\,F^a_{\mu\nu},\,
\tilde{{\Scr F}}_{i\mu\nu}, \,\tilde{F}_{i\mu\nu},\,
\tilde{{\Scr H}}^a_{\mu\nu},\,\tilde{F}_{a\mu\nu}\} \,,
\end{equation}
and in table 16 we give their $O(1,1)^3$ gradings and the corresponding
$E_{7(7)}$ weights.

\begin{table}[ht]
\vskip 0.5 cm
\caption{Field strengths, $O(1,1)^3$ gradings, and corresponding weights.}
\begin{center}
\vskip 0.5 cm
\begin{tabular}{|c|c|c|}\hline
Sp--section & $O(1,1)^3$--grading & weight 
\\\hline
${\Scr F}^i_{\mu\nu}$ & $(-1,-1,-{1\over 2})$& $-\epsilon_i-
{1\over\sqrt{2}}\epsilon_7$ 
\\\hline
$F^i_{\mu\nu}$ &$(1,-1,-{1\over 2})$& $w+\epsilon_j+\epsilon_k$   
\\\hline
${\Scr H}_{a\mu\nu}$  & $(-1,0,-{1\over 2})$& $\epsilon_a-
{1\over\sqrt{2}}\epsilon_7$  
\\\hline
$F^a_{\mu\nu}$  & $(-1,0,{1\over 2})$ & $w+\epsilon_b+\epsilon_c$
\\\hline
$\tilde{{\Scr F}}_{i\mu\nu}$ & $(1,1,{1\over 2})$& $\epsilon_i+
{1\over\sqrt{2}} \epsilon_7$ 
\\\hline
$\tilde{F}^i_{\mu\nu}$ &$(-1,1,{1\over 2})$& $-w-\epsilon_j-\epsilon_k$
\\\hline
$\tilde{{\Scr H}}^a_{\mu\nu}$  & $(1,0,{1\over 2})$ & $-\epsilon_a+
{1\over\sqrt{2}}\epsilon_7$  
\\\hline
$\tilde{F}_{a\mu\nu}$  & $(1,0,-{1\over 2})$& $-w-\epsilon_b-\epsilon_c$ 
\\\hline
\end{tabular}
\vskip 0.5 cm
\end{center}
\end{table}

The action of an infinitesimal duality transformation $\xi_{ij}\,T^{ij}+
\xi^{a}_i\,T_{a}^i+\xi_{ia}\,T^{ia}+\xi^{ab}\,T_{ab}+\xi\, T$ on the
symplectic section is
\begin{eqnarray}
\delta {\Scr F}^i &=& 0\,,
\nonumber \\
\delta F^i &=& \xi\,{\Scr F}^i \,,
\nonumber \\
\delta {\Scr H}_{a} &=&\xi_{ia}\, {\Scr F}^i\,,
\nonumber \\
\delta F^a &=& \xi_{ab}\,{\Scr H}_b+\xi^a_i\,{\Scr F}^i \,,
\nonumber\\
\delta \tilde{{\Scr F}}_i &=& -\xi_{ia}\, \tilde{{\Scr H}}^{a}-\xi^a_i\,
 \tilde{F}_a-2\,\xi_{ij}\, F^j-\xi\, \tilde{F}_i\,,
\nonumber \\
\delta \tilde{F}_i &=& \xi_{ia}\,F^a+\xi^a_i\, {\Scr H}_{a}+
2\,\xi_{ij}\,{\Scr F}^j\,,
\nonumber\\
\delta \tilde{{\Scr H}}^{a} &=& \xi^{ab}\,\tilde{F}_a+\xi^a_i\,F^i+\xi\, F^a\,,
\nonumber \\
\delta\tilde{F}_a &=& \xi_{ia}\,F^i+\xi\, {\Scr H}_a \,.
\end{eqnarray}
The explicit symplectic realisation of the $N_6$ generators
together with the computation of the vector kinetic matrix can be
found in appendix. 

\subsection{$T_5\times T_1$ IIA orientifold with $D8$--branes} 

\noindent{\it Solvable algebra of global symmetries.\ \ } The embedding of
the $sl(2,\mathbb{R})+ so(6,6)$ algebra inside $e_{7(7)}$ is defined by
the following identification of the simple roots
\begin{eqnarray}
\beta_1 &=& \epsilon_1-\epsilon_2\,,
\nonumber \\
\beta_2 &=& \epsilon_2-\epsilon_3\,,
\nonumber \\
\beta_3 &=& \epsilon_3-\epsilon_4\,,
\nonumber \\
\beta_4 &=& \epsilon_4-\epsilon_5\,,
\nonumber \\
\beta_5 &=& \epsilon_5+\epsilon_6\,,
\nonumber \\
\beta_6 &=& a+\epsilon_5\,,
\end{eqnarray}
for the $so(6,6)$ factor, and
\begin{equation}
\beta  = a+\epsilon_1+\epsilon_2+\epsilon_3+\epsilon_4+\epsilon_5\,,
\end{equation}
for the $sl(2,\mathbb{R})$ one. The correspondence axion--root is quite simple
and is summarised in table 17.

\begin{table}[ht]
\vskip 0.5 cm
\caption{Axionic fields for the $T_5 \times T_1$ IIA orientifold, generators 
of $Solv (so(6,6))$, $O(1,1)^3$ gradings, and $GL(5,\mathbb{R})$ 
representations.}
\begin{center}
\vskip 0.5 cm
\begin{tabular}{|c|c|c|c|c|}\hline
$ GL(5)$--rep. & generator & root & field & dim. 
\\\hline
--- & $T_{(0,0,0)}$ & $\{\epsilon_i-\epsilon_j,\,\epsilon_a-\epsilon_b\}$ 
{\scriptsize ($i<j $)} & $\{g_{ij}\}$ & 10 
\\\hline
${\bf 5}_{(0,0,1)}$ & $T^{i}$ &$a+\epsilon_i$ & $C_i$ & 5  
\\\hline
${\bf 5}_{(0,1,0)}$  & $T^{\prime i}$ & $\epsilon_i+\epsilon_6 $& $B_{i6}\,
\equiv\, B_i$ & 5 
\\\hline
${\bf 10}_{(0,1,1)}$ & $T^{ij}$ &$ a +\epsilon_i+\epsilon_j+\epsilon_6 $ &
$C_{ij6}\,\equiv\, C_{ij}$ & 10
\\\hline\hline
${\bf 1}_{(2,0,0)}$ & $T$ & $\beta$ &$C_{\mu\nu 6}\,\equiv\, c $ & 1 
\\\hline
\end{tabular}
\vskip 0.5 cm
\end{center}
\end{table}

The triple grading refers to three $O(1,1)$ groups generated by $H_{\beta},
\,H_{\lambda^5},\,H_{\lambda^6}$, all commuting
with $SL(5,\mathbb{R})$,  and parametrised by the  moduli
$\beta\cdot h,\,h_5,\,h_6$:
\begin{eqnarray}
O(1,1)_0  &\rightarrow & e^{\beta\cdot h}\,=\,V_5\, e^{\frac{\phi}{4}} \,,
\nonumber\\
O(1,1)_1 &\rightarrow& e^{h_5}\,=\,(V_5)^{\frac{1}{5}}\,
V_1\,e^{\frac{\phi}{2}},
\nonumber\\
O(1,1)_2 &\rightarrow & e^{h_6}\,=\,(V_5)^{\frac{1}{5}}\,e^{-\frac{3}{4}\,
\phi} \,.
\end{eqnarray}
The generators $T^{\prime i}$, $T^i$ and $T^{ij}$ form now a 
twenty--dimensional 
solvable subalgebra $N_8$ of $Solv(so(6,6))$:
\begin{equation}
N_8 = B_{i6}\,T^{\prime i}+C_i\,T^{i}+C_{ij}\, T^{ij} \,,
\end{equation}
whose algebraic structure is encoded in the non--vanishing commutator
\begin{equation}
\label{nncom1}\left[T^i,\,T^{\prime j}\right] = T^{ij} \,.
\end{equation}
A possible choice for the the coset representative is then
\begin{equation}
L = e^{C_{ij}\,T^{ij}}\,e^{B_{i6}\,T^{\prime
i}}\,e^{C_{i}\,T^{i}}\,e^{c\,T}\,\mathbb{E} \,, \label{coset5t1} 
\end{equation}
with the $\mathbb{E}$ parameterising the submanifold:
\begin{equation}
O(1,1)_0\times O(1,1)_1\times O(1,1)_2\times\frac{SL(5,\mathbb{R})}{SO(5)} \,.
\end{equation}
Under an infinitesimal transformation $\xi_{ij}\,T^{ij}+\xi_{i}\,T^i+
\xi_{i}^\prime\,T^{\prime i}$ of $N_8$ the variation of the axionic scalars is
\begin{eqnarray}
\delta C_{ij} &=& \xi_{[i}\, B_{j]6}+\xi_{ij} \,,
\nonumber\\
\delta B_{i6} &=& \xi^\prime_{i} \,,
\nonumber\\
\delta C_{i} &=& \xi_{i} \,.
\end{eqnarray}

\noindent{\it Vector fields.\ \ } The vector fields of this model are
$G^i_\mu$, $C_{i6\mu}$, $C_{\mu}$, $B_{a\mu}$, and we name the corresponding
field strengths ${\Scr F}^i_{\mu\nu}$, $F_{i6\mu\nu}$, $F_{\mu\nu}$,
${\Scr H}_{6\mu\nu}$. The symplectic section of the field strengths and their 
duals is
\begin{equation}
\{ {\Scr F}^i_{\mu\nu} ,\, F_{i6\mu\nu},\, F_{\mu\nu}, \, {\Scr H}_{6\mu\nu},
\, \tilde{\Scr F}_{i\mu\nu} ,\, \tilde{F}^{i6}_{\mu\nu},\, 
\tilde{F}_{\mu\nu}, \, \tilde{\Scr H}^6_{\mu\nu} \} \,,
\end{equation}
and in table 18 we give their $O(1,1)^3$ gradings and the corresponding
$E_{7(7)}$ weights.

\begin{table}[ht]
\vskip 0.5 cm
\caption{Field strengths, $O(1,1)^3$ gradings, and corresponding weights.}
\begin{center}
\vskip 0.5 cm
\begin{tabular}{|c|c|c|}\hline
Sp--section & $O(1,1)^3$--grading & weight 
\\\hline
${\Scr F}^i_{\mu\nu}$ & $(-1,-1,-{1\over 2})$& $-\epsilon_i-
{1\over\sqrt{2}} \epsilon_7$ 
\\\hline
$F_{i6\mu\nu}$ &$(-1,1,{1\over 2})$& $w+\epsilon_i+\epsilon_6$   
\\\hline
$F_{\mu\nu}$  & $(-1,-1,{1\over 2})$& $w$  
\\\hline
${\Scr H}_{6\mu\nu}$  & $(-1,1,-{1\over 2})$ & $\epsilon_6-
{1\over\sqrt{2}} \epsilon_7$
\\\hline
$\tilde{{\Scr F}}_{i\mu\nu}$ & $(1,1,{1\over 2})$& $\epsilon_i+
{1\over\sqrt{2}}\epsilon_7$ 
\\\hline
$\tilde{F}^{i6}_{\mu\nu}$ &$(1,-1,-{1\over 2})$& $-w-\epsilon_i-\epsilon_6$
\\\hline
$\tilde{F}_{\mu\nu}$  & $(1,1,-{1\over 2})$& $-w$ 
\\\hline
$\tilde{{\Scr H}}^6_{\mu\nu}$  & $(1,-1,{1\over 2})$ & $-\epsilon_6+
{1\over\sqrt{2}} \epsilon_7$
\\\hline
\end{tabular}
\vskip 0.5 cm
\end{center}
\end{table}

The action of an infinitesimal transformation $\xi_{ij}\,T^{ij}+\xi_{i}\,T^i
+\xi_{i}^\prime\,T^{\prime i}+\xi\, T$ on the symplectic section is
\begin{eqnarray}
\delta{\Scr F}^i &=& 0\,,
\nonumber \\
\delta F_{i6} &=& \xi_i\,{\Scr H}_6-\xi^\prime_i\, F+\xi_{ij}\,{\Scr F}^j\,,
\nonumber \\
\delta F &=& -\xi_i\,{\Scr F}^i\,,
\nonumber \\
\delta{\Scr H}_6 &=& \xi^\prime_i\, {\Scr F}^i \,,
\nonumber \\
\delta \tilde{{\Scr F}}_i &=& \xi_i\,\tilde{F}-\xi^\prime_i\, {\Scr H}^6+
\xi_{ij}\,\tilde{F}^{j6}+\xi\,F_{i6}\,,
\nonumber \\ 
\delta \tilde{F}^{i6} &=& \xi\, {\Scr F}^i\,,
\nonumber \\
\delta\tilde{F} &=& \xi^\prime_i\,\tilde{F}^{i6}+\xi\,{\Scr H}_6 \,,
\nonumber\\
\delta \tilde{{\Scr H}}^6 &=& -\xi_i\,\tilde{F}^{i6}+\xi\, F \,.
\end{eqnarray}
The explicit symplectic realisation of the $N_8$ generators, together with the 
computation of the vector kinetic matrix can be found in appendix.

\section{Fluxes and gauged supergravity: local Peccei--Quinn symmetry as 
gauged duality transformations.}

In the present section we consider the deformation of ${\Scr N}=4$
supergravity induced by the presence of fluxes. We shall restrict
our analysis, here, only to IIB orientifolds with some (three--form) fluxes 
turned on, while we shall defer the study of more general fluxes and of the 
gauge structure of other models elsewhere.

Differently to what happened in the well--studied $T_6 / \mathbb{Z}_2$ 
orientifolds, non--abelian gauged supergravities (for the bulk sector) now
emerge, due to the presence of gauge fields originating from the 
ten--dimensional metric, and of axionic scalars associated to the NS--NS 
two--form $B$.

\subsection{The $T_4\times T_2$ IIB orientifold model}

In this model, the allowed three--form fluxes are
$H^{\lambda}_{ij}=\{H_{aij},\, F_{aij}\}$, and are in
correspondence with the representation ${\bf (4,6)}_{+2}$ of
$SO(2,2)\times GL(4,\mathbb{R})$. The grading simply counts the
number of indices along the internal $T_4$ and, more specifically,
is associated to the subgroup $O(1,1)_1 \subset GL(4,\mathbb{R})$.
As mentioned in the introduction, inspection of the dimensionally
reduced three--form kinetic term indicates for the
four--dimensional theory a gauge group ${\Scr G}_g$ with
connection $\varOmega_g=X_i\, G^i_\mu+X_\lambda A^\lambda_\mu$ and
the following structure:
\begin{equation}
\left[X_i,\,X_j\right] = H_{ij}^\lambda X_\lambda \,.
\label{galg}
\end{equation} 
We may identify the gauge generators with isometries as follows:
\begin{eqnarray}
X_{i} &=& -H^{\lambda}_{ij}\,T^j_\lambda\,,
\nonumber \\
X^\lambda &=& {\textstyle\frac{1}{2}} \,H^{\lambda^\prime
}_{ij}\,T^{ij} \,. \label{isogen}
\end{eqnarray} 
Using relations (\ref{commutator}) and the property
\begin{equation}
H^\lambda_{k[j} H_{i]\ell\lambda } = 
{\textstyle \frac{1}{2}} \, H^\lambda_{ij} H_{k\ell\lambda }-
{\textstyle\frac{1}{4}}\, H\,\epsilon_{ijkl}\,,
\end{equation}
where $H=H_{ ij}^\lambda\,H_{\lambda}^{ij}$, one can show that the generators 
defined in (\ref{isogen}) fulfil the following algebraic relations
 \begin{equation}
\left[ X_i,\,X_j\right] = H^{\lambda}_{ij}\,X_\lambda-
{\textstyle\frac{1}{4}}H\, T_{ij} \,, \label{algebra}
\end{equation}
which coincide with (\ref{galg}) only if $H\,=\,0$ which amounts
to the condition that $\int_{T_6}\, F_{(3)}\wedge H_{(3)}\,=\,0$ 
(this condition is consistent with a constraint found in \cite{toappear} 
on the embedding matrix of a new gauge group in the ${\mathcal N}\,=\,8$ 
theory, which seems to yield an ${\mathcal N}\,=\, 8$ ``lifting'' of the 
type IIB orientifold models $T_{p-3}\times T_{9-p}$ discussed here).
Under this condition the gauge group is indeed contained in the
isometry group of the scalar manifold. Moreover it can be verified
that under the duality action of the gauge generators defined in
(\ref{isogen}) the vector fields transform in the co--adjoint of
the gauge group ${\Scr G}_g$ and  thus provide a consistent
definition for the gauge connection $\varOmega_g$ . The variation
of the gauge potentials under an infinitesimal transformation with
parameters $\xi^\lambda,\,\xi^i$ reads
\begin{eqnarray}
\delta A^{{\lambda}}_\mu &=& \xi^i\,H_{ij}^{{\lambda}}\,G^j_\mu+
\partial_\mu \xi^\lambda \,,
\nonumber\\
\delta G^i_\mu &=&\partial_\mu \xi^i \,,
\nonumber\\
\delta C_{ijk\mu}&=&0 \,,
\end{eqnarray}
and is compatible with the following non--abelian field strengths
\begin{eqnarray}
F^\lambda_{\mu\nu} & = &\partial_\mu A^\lambda_\nu-\partial_\nu
A^\lambda_\mu -H_{ij}^\lambda\,G^i_\mu\,G^j_\nu \,,
\nonumber\\
{\Scr F}^i_{\mu\nu} &=&\partial_\mu G^i_\nu-\partial_\nu G^i_\mu \,,
\nonumber\\
F^i_{\mu\nu} &=& \epsilon^{ijkl}\,\left(\partial_\mu
C_{jkl\nu}-\partial_\nu C_{jkl\mu}\right) \,.
\end{eqnarray}
The $C_{ij}$ and $\varPhi^\lambda_i$ scalars are also charged and, up to 
rotations, subject to shifts
\begin{eqnarray}
\delta C_{ij} &=&
{\textstyle\frac{1}{2}}\,H_{ij}^\lambda\,\xi_\lambda-
{\textstyle\frac{1}{2}} \,\xi^k\,
H^{\lambda}_{k[i}\,
\Phi_{j]\lambda} ,
\nonumber\\
\delta \Phi^\lambda_i &=& H_{ij}^\lambda\,\xi^j \,,
\end{eqnarray}
and their kinetic terms are modified accordingly by covariantisations
\begin{eqnarray}
D_\mu C_{ij} &=& \partial_\mu
C_{ij}-{\textstyle\frac{1}{2}}\,H_{ij\,\lambda}\,A^{{\lambda}}_{\mu}+
{\textstyle\frac{1}{2}} \,G_\mu^k\,
H^{\lambda}_{k[i}\,
\Phi_{j]\lambda},
\nonumber\\
D_\mu\Phi^\lambda_i &=& \partial_\mu \Phi^\lambda_i
-H^\lambda_{ij}\,G_\mu^j \,.
\end{eqnarray}

\noindent{\it Chern--Simons terms.\ \ } The gauge group 
consists of Peccei--Quinn transformations that shift the real part of the
vector kinetic matrix ${\Scr N}$ (the generalised theta angle). In 
\cite{deWit:1984px},\cite{deWit:1986pj}, 
it was shown that such a local transformation is a symmetry of the Lagrangian 
provided suitable generalised Chern--Simons terms are introduced. 

In the case at hand, the new contribution to the Lagrangian is
\begin{equation}
{\Scr L}_{\rm c.s.} \propto 
\epsilon^{\mu\nu\rho\sigma}\,\left( \,H_{\lambda\,i
j^\prime}\,A^\lambda_\mu\,G^i_\nu\,\partial_\rho
C^{j^\prime}_\sigma
+{\textstyle\frac{1}{8}} \,H_{\lambda\,i
j^\prime}\,H^\lambda_{k\ell}\,G^i_\mu\,C^{j^\prime}_\nu\,G^k_\rho\,
G^\ell_\sigma\right) \,, 
\end{equation}
corresponding to the non--vanishing entries
\begin{equation}
C_{\lambda,\,i j^\prime} =  -H_{\lambda\,i j^\prime}\quad {\rm and}
\qquad C_{i,\,\lambda j^\prime}\,=\,H_{\lambda\,i j^\prime}
\end{equation}
where, in general, the coefficients $C_{\varGamma , \varLambda \varSigma}$
define the moduli--independent gauge variation of the real part of the kinetic
matrix ${\Scr N}$
\begin{equation}
\delta_\xi \, {\rm Re}\, {\Scr N} _{\varLambda \varSigma} = 
\xi^\varGamma C_{\varGamma ,\varLambda\varSigma}\,.
\end{equation}

\subsection{Type $T_2\times T_4$ IIB orientifold model.}
Let us consider the $T_2\times T_4$ model in presence of the fluxes 
$H_{ija}=\epsilon_{ij}\, H_a$ and $F_{iab}$. These fluxes appear as
 structure constants 
\begin{eqnarray}
\left[X_i,\,X_j\right] &=& \epsilon_{ij}\, H_a\, X^a\,,
\nonumber \\
\left[X_i,\,X^a\right] &=& F_i^{ab}\, X_b \,, \label{algebra2}
\end{eqnarray}
of the  gauge algebra ${\Scr G}_g\,\equiv\,\{X_i,\,X^a,\,X_a\}$ with connection
$\varOmega^g_\mu = G^i_\mu \, X_i+B_{a\mu}\, X^a+ C^a_\mu\,X_a$, all other
commutators vanishing.

The identification
\begin{eqnarray}
\label{gauge2} X^\prime_i &=&-F_i^{ab}\, T_{ab}+H_a\, T^a_i \,,
\nonumber \\
X^{a\prime } &=& F_i^{ab}\, T^i_b \,,\,\,\,X_a^\prime\,=\,-H_a\, T\,,
\end{eqnarray}
of the gauge generators with the isometries of the solvable algebra $N_5$, 
reproduces only a contracted version of the algebra (\ref{algebra2}) in 
which three of the central charges $X_a$ vanish and we are left with 
$X^\prime_a\,=\,- H_a\, T$. If we denote by $\{X^\perp\}=\, \{X_a\}/ 
\{X^\prime_a\}$ these three central generators, we see that the subgroup 
${ \Scr G}^\prime_g\,=\,\{X^\prime_i,\,X^{a\prime },\,X^\prime_a\}$ of the 
isometry group
 which is gauged coincides with the quotient:
\begin{equation}
 {\Scr G}^\prime_g\,\equiv \,{\Scr G}_g/\{X^\perp\} \,,
\end{equation}
that amounts to imposing the vanishing of the central terms on all fields.

On the other hand, transformations generated by the operators in (\ref{gauge2})
induce isometry transformations with parameters:
\begin{eqnarray}
\xi_{ia} &=& -\xi_i\, H_a\,,
\nonumber \\
\xi^{ab} &=& -\xi^i\, F_i^{ab}\,,
\nonumber \\
\xi^{a}_i &=& \xi_b\, F_i^{ba} \,, 
\nonumber \\
\xi &=& - H_a \xi^a \,, \label{gisom}
\end{eqnarray}
where $\xi_i\,=\,\epsilon_{ij}\,\xi^j$. Using eqs. (\ref{isom})
and (\ref{gisom}), one can then verify that the vectors
$G^i_\mu$, $B_{a\mu}$  and $C^a_\mu$ transform in the co--adjoint
representation of ${\Scr G}_g$ under the duality action
generated by  $\{X_i,\,X^a,\,X_a\}$, so that the above definition of
the gauge connection $\varOmega^g_\mu$ is consistent:
\begin{eqnarray}
\delta B_{a\mu} &=& \xi^i\, G^j_\mu\,\epsilon_{ij}\, H_a+\partial_\mu 
\xi_a\,=\,-\xi_i\, G^i_\mu\, H_a+\partial_\mu \xi_a \,,
\nonumber\\
\delta C^a_\mu &=&\xi^i\,B_{b\mu}\,F_i^{ba}-G^i_\mu\,\xi_b\, F_i^{ba}+
\partial_\mu \xi^a \,,
\nonumber\\
\delta G^i_\mu &=&\partial_\mu \xi^i \,.
\end{eqnarray}
Notice that the action of the central charges $X_a$ amounts just to
a gauge transformation on $C^a_\mu$. These ten vectors can
therefore be used to gauge the group ${\Scr G}_g$, and the
non--abelian field strengths read
\begin{eqnarray}
{\Scr H}_{a\mu\nu} &=& \partial_\mu B_{a\nu}-\partial_\nu B_{a\mu}-
\epsilon_{ij}\,H_a\, G^i_\mu\,G^j_\nu \,,
\nonumber\\
F^a_{\mu\nu}&=& \partial_\mu C^a_{\nu}-\partial_\nu C^a_{\mu}+F_i^{ab}\, 
G^i_\mu\,B_{b\nu}-F_i^{ab}\, G^i_\nu\,B_{b\mu} \,,
\nonumber\\
{\Scr F}^i_{\mu\nu}&=&\partial_\mu G^i_{\nu}-\partial_\nu G^i_{\mu} \,.
\end{eqnarray}
Since ${\Scr G}_g$ is not part of the global symmetries
of the Lagrangian, we should restrict ourselves to the quotient $G_g$,
{\it i.e.} we demand that central charges $\{T,\,X_a\}$ vanish on all 
physical fields. The gauge transformations of the scalar fields
\begin{eqnarray}
\delta c &=& -H_a\, \xi^a+\xi_{a}\,
F_i^{ab}\,B_{jb}\,\epsilon^{ij} \,,
\nonumber\\
\delta C^a_i &=& \xi_b\,  F_i^{ba}+\xi^j\,F_j^{ab}\, B_{bi}\,,
\nonumber\\
\delta B_{ia}&=&-\xi_i\, H_a \,,
\nonumber\\
\delta C_{ab}&=&-\xi^i\,F_{iab} \,,
\end{eqnarray}
are then compatible with the covariant derivatives
\begin{eqnarray}
D_\mu c&=&\partial_\mu c+H_a\,
C^a_\mu -B_{a\mu}\,
F_i^{ab}\,B_{jb}\,\epsilon^{ij} \,,
\nonumber\\
D_\mu C^a_i&=&\partial_\mu C^a_i-B_{b\mu}\, 
F_i^{ba}-G^j_\mu\,F_j^{ab}\, B_{bi} \,,
\nonumber\\
D_\mu B_{ia}&=&\partial_\mu B_{ia}+G_{i\mu}\,H_a \,,\nonumber\\
D_\mu C_{ab}&=&\partial_\mu C_{ab}+G_\mu^i\, F_{iab}\,.
\end{eqnarray}

\noindent{\it Chern--Simons terms.\ \ } Also in this case local Peccei--Quinn
transformations demand the inclusion in the Lagrangian of the Chern--Simons
terms
\begin{eqnarray}
{\Scr L}_{\rm c.s.} &=& \epsilon^{\mu\nu\rho\sigma}\,\Bigl( H_a\,G^i_\mu\,
C_{i\nu}\,\partial_\rho C^a_\sigma-H_a\, C^a_\mu\,C_{i\nu}\,\partial_\rho
G^i_\sigma
-\epsilon^{ij}\,F_j^{ab}\,B_{a\mu}\,C_{i\nu}\,\partial_\rho
B_{b\sigma}
\nonumber\\
&& + {\textstyle\frac{1}{8}}\,H_a\,F_k^{ab}\,G^i_\mu\,C_{i\nu}\,G^k_\rho\,
B_{b\sigma} - {\textstyle\frac{1}{8}}\,\epsilon^{ij}\,H_a\,F_{j}^{ab}\,
B_{b\mu}\,C_{i\nu}\,G^k_\rho\,G_{k\sigma} \Bigr) \,,
\end{eqnarray}
corresponding to the non--vanishing components
\begin{eqnarray}
C_{i,}{}^j{}_{ a} &=& \delta^j_i \,H_a\,,
\nonumber \\
C_{a,\,i}{}^j &=& -\delta^j_i\,H_a\,,
\nonumber \\
C^{a,\,i b} &=& - \epsilon^{ij}\,F_j^{ab} \,,
\end{eqnarray}
of the $C_{\varGamma, \varLambda\varSigma}$ coefficients.

\section{Conclusions and outlooks}

In the present paper, we have investigated the symmetries and the structure of
several $T_6$ orientifolds which, in absence of fluxes, have ${\Scr N} =4$
supersymmetries in four dimensions. we have not addressed here the question 
of vacua with some residual supersymmetry, that will be the subject of future 
investigations. All these models lead to different
low--energy supergravity descriptions. When fluxes are turned on, the deformed 
Lagrangian is described by a gauged ${\Scr N}=4$ supergravity and fermionic 
mass--terms and a scalar potential are developed.

The low--energy Lagrangians underlying these orientifolds are different 
versions of gauged ${\Scr N} =4$ supergravity with six bulk vector multiplets 
and additional Yang--Mills multiplets living on the brane world--volume.
The gaugings  are based on quotients (with respect to some central charges) 
of nilpotent subalgebras of $so(6,6)$. These nilpotent subalgebras are 
basically generated by the axion symmetries associated to R--R scalars 
and to NS--NS scalars originating  from the two--form $B$--field.

Along similar lines, one can also consider new examples of orientifolds with
${\Scr N} =2,1$ four--dimensional supersymmetries, with and/or without fluxes.

\vskip 1cm

\noindent
{\bf Acknowledgements} 
M.T. would like to thank H. Samtleben for useful discussions and 
the Th. Division of CERN, where part of this work has
 been done, for their kind hospitality.
The work of S.F. has been supported in part by European Community's Human 
Potential Program under contract HPRN-CT-2000-00131 Quantum Space-Time, in 
association with INFN Frascati National Laboratories and by D.O.E. grant 
DE-FG03-91ER40662, Task C. The work of M.T. is supported by a European
 Community Marie Curie Fellowship under contract HPRN-CT-2001-01276.

\appendix
\section*{Appendix. Symplectic realisation of the solvable generators}
\label{appendiceA}

In this appendix, we give the coset representatives of our models in the 
symplectic basis of vector fields. This is needed in order to compute the 
kinetic matrix ${\Scr N}_{\varLambda\varSigma}$, which is a complex symmetric
matrix in the space of vectors in the theory. Its imaginary and real parts
describe the terms
\begin{equation}
{\rm Im } \, {\Scr N}_{\varLambda\varSigma} \, 
F_{\mu\nu}^\varLambda F^{\varSigma
\, \mu\nu} + {\textstyle{1\over 2}}\, 
{\rm Re} \, {\Scr N}_{\varLambda\varSigma} \, \epsilon^{\mu\nu\rho\sigma}
F^\varLambda_{\mu\nu} F^\varSigma_{\rho\sigma} \,.
\end{equation}

\paragraph{Model $T_4\times T_2$} The $Sp(24,\mathbb{R})$ representation of 
the solvable generators in model 1 in the basis (\ref{symba}) is:
\begin{eqnarray}
T &=& \left(
\matrix{0 & 0 & 0 &0 & 0 & 0 
\cr 
0 & 0 & 0 &0 & 0 & 0 
\cr 
0 & \bfone & 0 &0 & 0 & 0 
\cr 
\eta  & 0 &0 & 0 & 0 &0
\cr  
0 & 0 & 0 &0 & 0 & -\bfone 
\cr 
0 & 0 & 0 &0 & 0 & 0  }
\right) \,,
\nonumber\\
T &=& \left(\matrix{M^T & 0 & 0 &0 & 0 & 0 
\cr 
0 & 0 & 0 &0 & 0 & 0 
\cr 
0 & 0& 0 &0 & 0 & 0 
\cr 
0& 0 &0 & -M & 0 &0 
\cr  
0 & 0 & 0 &0 & 0 & 0 
\cr 
0 & 0 & 0 &0 & 0 & 0  }
\right)\,, 
\nonumber\\
T^i_\lambda &=& \left(\matrix{0 & -(t^i_\lambda)^T & 0 &0 & 0 & 0
\cr 
0 & 0 & 0 &0 & 0 & 0 
\cr 
0 & 0&0 &0 & 0 & 0 
\cr 
0& 0 &  -(t^i_\lambda\,\eta)^T & 0& 0 &0
\cr 
0 & 0 & 0 & t^i_\lambda & 0 & 0
\cr
-t^i_\lambda\,\eta & 0 & 0 &0 & 0 & 0  }
\right) \,,
\nonumber\\
T^{ij} &=&  \left(\matrix{0 & 0 & 0 &0 & 0 & 0 
\cr 
0 & 0 & 0 &0 & 0 & 0 
\cr 
0 & 0&0 &0 & 0 & 0 
\cr 
0& 0 & 0 & 0& 0 &0
\cr 
0 & 0 & -t^{ij} &0 & 0 & 0
\cr  
0 & t^{ij} & 0 &0 & 0 & 0  }
\right) \,,
\end{eqnarray}
where each block is a  $4\times 4$ matrix, $\bfone$ denotes the  identity 
matrix, $\eta \equiv \eta_{\lambda\lambda^\prime}$ and
\begin{equation}
(t^i_\lambda)_j{}^{\lambda^\prime} = \delta_j^i\,
\delta_\lambda^{\lambda^\prime}\,, \qquad (t^{ij})_{kl}\,=\, 
\delta^{i}_{k}\delta^j_l-\delta^{i}_{l}\delta^j_k \,.
\end{equation}
The coset representative is
\begin{equation}
L = \exp (C_{ij} T^{ij} ) \, \exp (\varPhi^\lambda_i T^i_\lambda )\,
\exp (c T) \, \mathbb{E} = 
\left(\matrix{{\sf A}& 0\cr {\sf C} & {\sf D}}\right) \,,
\end{equation}
where $\mathbb{E}$ parametrises the manifold
\begin{equation}
\mathbb{E} \in  O(1,1)_0\times \frac{SO(2,2)}{SO(2)\times SO(2)}\times 
\frac{GL(4,\mathbb{R})}{SO(4)} \,,
\end{equation}
and can be written in the following general form:
\begin{equation}
\mathbb{E} = \mbox{{\scriptsize $\left(\matrix{
e^{-\varphi}\,E_{(\ell)} & 0 & 0 &0 & 0 & 0 
\cr 
0 & e^{-\varphi}\,E & 0 &0 & 0 & 0 
\cr 
0 & 0 & e^{\varphi}\,E &0 & 0 & 0 
\cr 
0  & 0 &0 & e^{\varphi}\,\eta E_{(\ell)}\eta & 0 &0
\cr  
0 & 0 & 0 &0 &e^{\varphi}\, E^{-1} & 0 
\cr 
0 & 0 & 0 &0 & 0 & e^{-\varphi}\, E^{-1}  }\right)$}} \,,
\end{equation}
with
\begin{eqnarray}
E_{(\ell)}{}^\lambda{}_{\hat{\sigma}} &\in & \frac{SO(2,2)}{SO(2)\times 
SO(2)}\,, 
\nonumber \\
E^i{}_{\hat{\jmath}} &\in &  \frac{GL(4,\mathbb{R})}{SO(4)}\,,
\nonumber \\
e^{\varphi\,H} &\in& O(1,1)_0 \,,
\end{eqnarray}
the hatted indices being the rigid ones transforming under the isotropy group.
The blocks is $L$ read
\begin{eqnarray}
{\sf A} &=& \mbox{{\scriptsize
$\left(\matrix{e^{-\varphi}\,E_{(\ell)}{}^\lambda
{}_{\hat{\sigma}} & -e^{-\varphi}\,\varPhi_i^\lambda\, E^i{}_{\hat{\jmath}}
& 0\cr 0 & e^{-\varphi}\,E^i{}_{\hat{\jmath}} & 0\cr 0 &
c\,e^{-\varphi}\, E^i{}_{\hat{\jmath}} &
e^{\varphi}\,E^i{}_{\hat{\jmath}}}\right)$}} \,,
\nonumber\\ 
{\sf C}
&=&\mbox{{\scriptsize $\left(\matrix{c\,e^{-\varphi}\,
E_{(\ell)\lambda\hat{\sigma}} & -c\,e^{-\varphi}\, \varPhi_{\lambda
j}\, E^j{}_{\hat{\imath}} & -e^{\varphi}\, \varPhi_{\lambda j}\,
E^j{}_{\hat{\imath}}\cr c\,e^{-\varphi}\, \varPhi_{\delta
i}\,E_{(\ell)}{}^\delta {}_{\hat{\sigma}}
&-c\,e^{-\varphi}\,2\,\tilde{C}_{ij}E^j{}_{\hat{k}} &
-e^{\varphi}\,2\,
\tilde{C}_{ij} E^j{}_{\hat{k}}\cr -e^{-\varphi}\,
\varPhi_{\delta i}\,E_{(\ell)}{}^\delta {}_{\hat{\sigma}} &
e^{-\varphi}\,2\, \tilde{C}_{ij}E^j{}_{\hat{k}}& 0 }\right)$}} \,,
\nonumber\\
{\sf D}&=&\mbox{{\scriptsize
$\left(\matrix{e^{\varphi}\,E_{(\ell)\lambda} {}^{\hat{\sigma}} & 0
 & 0\cr e^{\varphi}\,\varPhi_i^\lambda E_{(\ell)\lambda}{}^{\hat{\sigma}} &
 e^{\varphi}\, E^{-1}{}_{i}{}^{\hat{\jmath}} & -c\,e^{-\varphi}\,
 E^{-1}{}_{i}{}^{\hat{\jmath}}\cr 0 & 0 & e^{-\varphi}\,
E^{-1}{}_{i}{}^{\hat{\jmath}}}\right)$}} \,,
\nonumber\\
\tilde{C}_{ij} &=& C_{ij}+{\textstyle\frac{1}{4}}\,
\varPhi_i^\lambda\,\varPhi_{\lambda j} \,.
\end{eqnarray}
In the sequel we shall need also the expression of ${\sf
A}^{-1}$:
\begin{equation}
{\sf A}^{-1} = \mbox{{\scriptsize $\left(\matrix{e^{\varphi}\,
E_{(\ell)}{}^{\hat{\sigma}}{}_\lambda & e^{\varphi}\,
E_{(\ell)}{}^{\hat{\sigma}}{}_\lambda \varPhi_i^\lambda &0\cr 0
&e^{\varphi}\,E^{-1}{}^{\hat{\imath}}{}_j &0\cr 0&
-e^{-\varphi}\,c\,E^{-1}{}^{\hat{\imath}}{}_j
&e^{-\varphi}\,E^{-1}{}^{\hat{\imath}}{}_j  }\right)$}} \,.
\end{equation}
In terms of the matrices $h,\,f$ 
\begin{equation}
f = {\textstyle\frac{1}{\sqrt{2}}}\,{\sf A}\,, \qquad h = 
{\textstyle\frac{1}{\sqrt{2}}}\,\left({\sf C}-{\rm
i}\, {\sf D} \right) \,, \label{fh} 
\end{equation}
the kinetic matrix is expressed as (see \cite{Andrianopoli:1996ve} and 
references therein) 
\begin{equation}
{\Scr N} =  h f^{-1}\,=\,
\left(\matrix{ N_{\lambda\lambda^\prime}&  N_{\lambda i}& 
N^\prime_{\lambda i}\cr *& N_{ij}& N^\prime_{ij}\cr *
&*& N^{\prime\prime}_{i
j} }\right)\,,\label{N} 
\end{equation}
and is  characterised by the following entries:
\begin{eqnarray}
N_{\lambda\lambda^\prime} &=& -{\rm
i}\,e^{2\varphi}\,E_{(\ell)\lambda}{}^{\hat{\sigma}}\,
E_{(\ell)\lambda^\prime}{}^{\hat{\sigma}}+c\,\eta_{\lambda\lambda^\prime}\,,
\nonumber \\
N_{\lambda
i} &=& -{\rm i}\,e^{2\varphi}\,E_{(\ell)\lambda}{}^{\hat{\sigma}}\,
E_{(\ell)\lambda^\prime}{}^{\hat{\sigma}}\,\varPhi^{\lambda^\prime}_i+c\,
\varPhi_{\lambda i}
\nonumber\\
N^\prime_{\lambda i} &=&-\varPhi_{\lambda i
}\,,
\nonumber \\
N_{ij} &=& -{\rm i} \left((e^{2\varphi}+e^{-2\varphi}\,c^2)\,
E^{-1}{}_i{}^{\hat{\imath}}E^{-1}{}_{j\hat{\imath}}+e^{2\varphi}\,
\varPhi^{\lambda}_i\,E_{(\ell)\lambda}{}^{\hat{\sigma}}\,
E_{(\ell)\lambda^\prime}{}^{\hat{\sigma}}\,\varPhi^{\lambda^\prime}_j\right)
+c\,\varPhi^{\lambda}_i\,
\varPhi_{\lambda j} \,,
\nonumber\\
N^\prime_{ij} &=& {\rm i}\,c \,\,e^{-2\varphi}
E^{-1}{}_i{}^{\hat{\imath}}E^{-1}{}_{j\hat{\imath}}-2\,\tilde{C}_{ij}\,,
\nonumber \\
N^{\prime\prime}_{i
j} &=& -{\rm i}\,e^{-2\varphi}\,
E^{-1}{}_i{}^{\hat{\imath}}E^{-1}{}_{j\hat{\imath}} \,.
\end{eqnarray}

\paragraph{Model $T_2\times T_4$} The $Sp(24,\mathbb{R})$ representation of 
the solvable generators in model 2 in the basis (\ref{symba2}) is:
\begin{eqnarray}
T^\prime &=&\left(\matrix{ 0&0&0&0&0&0&0&0\cr \epsilon_{ij}&0&0&0&0&0&0&0\cr
0&0&0&0&0&0&0&0\cr 0&0&0&0&0&0&0&0\cr 0&0&0&0&0&\epsilon_{ij}&0&0\cr 
0&0&0&0&0&0&0&0\cr 0&0&0&\bfone &0&0&0&0\cr 0&0&\bfone &0&0&0&0&0}\right) \,,
\nonumber\\
T_{ab}&=&\left(\matrix{ 0&0&0&0&0&0&0&0\cr 0&0&0&0&0&0&0&0\cr
0&0&0&0&0&0&0&0\cr 0&0&\delta^{cd}_{ab}&0&0&0&0&0\cr 0&0&0&0&0&0&0&0\cr 
0&0&0&0&0&0&0&0\cr 0&0&0&0 &0&0&0&\delta^{cd}_{ab}\cr 0&0&0 &0&0&0&0&0}
\right) \,,
\nonumber\\
T^{ia}&=&\left(\matrix{ 0&0&0&0&0&0&0&0\cr 0&0&0&0&0&0&0&0\cr
\delta^a_b\,\delta^i_j&0&0&0&0&0&0&0\cr 0&0&0&0 &0&0&0&0\cr 0&0&0&0&0&0&-
\delta^i_j\,\delta^a_b &0\cr 0&0&0&-\epsilon^{ij}\,\delta^a_b&0&0&0&0\cr 
0&0&0&0&0&0&0&0\cr 0&-\delta^a_b\,\epsilon^{ij}&0 &0&0&0&0&0}\right)\,,
\nonumber\\
T^{i}_a&=&\left(\matrix{ 0&0&0&0&0&0&0&0\cr 0&0&0&0&0&0&0&0\cr 
0&0&0&0 &0&0&0&0\cr
-\delta^b_a\,\delta^i_j&0&0&0&0&0&0&0\cr 0&0&0&0&0&0&0&\delta^i_j\,\delta^b_a
\cr 0&0&\epsilon^{ij}\,\delta^b_a &0&0&0&0&0\cr 
0&\delta^b_a\,\epsilon^{ij}&0 &0&0&0&0&0\cr 0&0&0&0&0&0&0&0}\right)\,,
\nonumber\\
T&=&\left(\matrix{ 0&0&0&0&0&0&0&0\cr 0&0&0&0&0&0&0&0\cr
0&0&0&0&0&0&0&0\cr 0&0&0&0&0&0&0&0\cr 0&\bfone &0&0&0&0&0&0\cr 
\bfone&0&0&0&0&0&0&0\cr 0&0&0&0&0&0&0&0\cr 0&0&0 &0&0&0&0&0}\right) \,.
\end{eqnarray}
The coset representative has the form:
\begin{equation}
L = e^{c^\prime\,T^\prime}\,e^{c\,T}\,e^{B_{ia}\,T^{ia}}\,e^{C_i^a\,T^i_a}
\,e^{C_{ab}\,T^{ab}}\,\mathbb{E}\,=\,\left(\matrix{{\sf A}& 0\cr {\sf C} 
&{\sf D} }\right) \,,
\end{equation}
where this time the matrix $\mathbb{E}$ describes the submanifold:
\begin{equation}
\mathbb{E} = O(1,1)_0\times \frac{GL(2,\mathbb{R})}{SO(2)}\times 
\frac{ GL(4,\mathbb{R})}{SO(4)} \,,
\end{equation}
and has the following form: 
\begin{equation}
\mathbb{E}=\mbox{\scriptsize{$\left(\matrix{e^{-\varphi}\,
E_2{}^i{}_{\hat{\jmath}} & 0&0&0&0&0&0&0\cr 0&e^{\varphi}\,
E_2^{-1}{}_i{}^{\hat{\jmath}} & 0&0&0&0&0&0\cr0& 0&e^{-\varphi}\,
E_4^{-1}{}_a{}^{\hat{b}} & 0&0&0&0&0\cr 0&0&0& e^{-\varphi}\,
E_4{}^a{}_{\hat{b}} & 0&0&0&0\cr 0&0&0&0& e^{\varphi}\,
E_2^{-1}{}_i{}^{\hat{\jmath}} &0&0&0\cr 0&0&0&0&0 &e^{-\varphi}\,
E_2{}^i{}_{\hat{\jmath}} &0&0\cr 0& 0&0&0&0&0& e^{\varphi}\,
E_4{}^a{}_{\hat{b}} &0\cr 0& 0&0&0&0&0&0& e^{\varphi}\,
E_4^{-1}{}_a{}^{\hat{b}} }\right)$}} \,,
\nonumber
\end{equation}
\begin{eqnarray}
E_2{}^i{}_{\hat{\jmath}}&\in
&\frac{GL(2,\mathbb{R})}{SO(2)}\,,
\nonumber \\
E_4{}^a{}_{\hat{b}} &\in&
\frac{GL(4,\mathbb{R})}{SO(4)}\,,
\nonumber \\
e^{\varphi\,H} &\in
& O(1,1)_0 \,.
\end{eqnarray}
The blocks ${\sf A},\,{\sf C},\,{\sf D}$ of $L$ can be conveniently described 
in terms of the following matrices $({\bf B})_{ia}=B_{ia}$, 
$({\bf C})_{i}{}^a= C_{i}^a$, $({\Scr C})^{ab}\,=\,C^{ab}$: 
{\small
\begin{eqnarray}
{\sf A}&=&\mbox{{\scriptsize $\left(\matrix{e^{-\varphi}\,E_2 &
0&0&0\cr e^{-\varphi}\,c^\prime\,\epsilon\, E_2 &e^{\varphi}\,
E_2^{-1}&0&0\cr e^{-\varphi}\,{\bf B}^t\, E_2 & 0&e^{-\varphi}\,
E_4^{-1}&0\cr -e^{-\varphi}\,{\bf C}^t\,E_2 &0&e^{-\varphi}\,{\Scr
C}\,E_4^{-1}&
e^{-\varphi}\,E_4}\right)$}} \,,
\nonumber\\
{\sf C}&=&\mbox{{\scriptsize
$\left(\matrix{e^{-\varphi}\,c^\prime\,(c\,\epsilon+{\bf B}\,{\bf
C}^t)\,E_2 & -e^{\varphi}\,(c\,\epsilon+{\bf B}\,{\bf
C}^t)\,\epsilon\,E_2^{-1} &e^{-\varphi}\,c^\prime\, ({\bf C}-{\bf
B}\,{\Scr C})\,E_4^{-1}&-e^{-\varphi}\,c^\prime\,{\bf B}\,E_4\cr
-e^{-\varphi}\,\epsilon\, (c\,\epsilon+{\bf B}\,{\bf C}^t)\,E_2 &
0&-e^{-\varphi}\,\epsilon\,({\bf C}-{\bf B}\,{\Scr
C})\,E_4^{-1}&e^{-\varphi}\,\epsilon\,{\bf B}\, E_4\cr
-e^{-\varphi}\,c^\prime \, {\bf C}^t\, E_2 &e^{\varphi}\,{\bf
C}^t\,\epsilon\, E_2^{-1}&e^{-\varphi}\,c^\prime \,{\Scr C}\,
E_4^{-1}&e^{-\varphi}\, c^\prime \,E_4\cr e^{-\varphi}\, c^\prime
\,{\bf B}^t\, E_2 &-e^{\varphi}\,{\bf B}^t\,\epsilon\, 
E_2^{-1}&e^{-\varphi}\,c^\prime \, E_4^{-1}&0}
\right)$}} \,,
\nonumber\\
{\sf D}&=&\mbox{{\scriptsize $\left(\matrix{e^{\varphi}\,E_2^{-1}&
e^{-\varphi}\,c^\prime\,\epsilon\, E_2& -e^{\varphi}\,{\bf B}\,
E_4 &e^{\varphi}\, ({\bf C}-{\bf B}\,{\Scr C})\, E_4^{-1}\cr
0&e^{-\varphi}\, E_2 & 0& 0\cr 0 & 0 &e^{\varphi}\, E_4
&e^{\varphi}\, {\Scr C}\, E_4^{-1}\cr
0&0&0&e^{\varphi}\,E_4^{-1}}\right)$}} \,,
\end{eqnarray}}
it is also useful to compute ${\sf A}^{-1}$:
\begin{equation}
{\sf A}^{-1}=\mbox{{\scriptsize
$\left(\matrix{e^{\varphi}\,E_2^{-1} & 0&0&0\cr
-e^{-\varphi}\,c^\prime\, E_2\,\epsilon &e^{-\varphi}\, E_2&0&0\cr
-e^{\varphi}\,E_4\,{\bf B}^t & 0&e^{\varphi}\, E_4&0\cr
e^{\varphi}\,E_4^{-1}\,({\Scr C}\,{\bf B}^t+{\bf C}^t)
&0&-e^{\varphi}\,E_4^{-1}\,{\Scr
C}&e^{\varphi}\,E_4^{-1}}\right)$}} \,.
\end{equation}
We then compute the kinetic matrix ${\Scr N}$ whose independent components are:
\begin{equation}
{\Scr N} = \left(\matrix{
N_{ij}& N_{i}{}^j & N_{i}{}^a & N_{ia}\cr *
& N^{ij} & N^{ia} & N^i{}_a \cr * & * & 
N^{ab} & N^a{}_b\cr * & * & * & N_{ab} }\right)\,,
\end{equation}
where
\begin{eqnarray}
N_{ij}&=&-{\rm
i}\,\left[E_2^{-1}{}_i{}^{\hat{\jmath}}\,E_2^{-1}{}_j{}^{\hat{\jmath}}\,
(e^{2\,\varphi}+ e^{-2\,\varphi}\, c^{\prime
2})+e^{2\,\varphi}\,B_{ia}\,E_4{}^a{}_{\hat{a}}\,E_4{}^b{}_{\hat{a}}\,B_{jb}+
\right.
\nonumber\\
&& \left. e^{2\,\varphi}\,(-B_{ic}\,
C^{ca}+C_i^a)\,E_4^{-1}{}_a{}^{\hat{a}}\,E_4^{-1}{}_b{}^{\hat{a}}
\,(C^{bd}\,B_{jd}+C_j^b)\right]-2\, B_{a(i}\,C^a_{j)}\, c^\prime \,,
\nonumber\\
{  N}_i{}^j &=& -{\rm i}\,e^{-2\,\varphi}\, c^\prime\,
\epsilon_{ik}\,E_2{}^k{}_{\hat{k}}\,
E_2{}^j{}_{\hat{k}}+c\,\delta_i{}^j-B_{ia}\,C^a_k\,\epsilon^{kj} \,,
\nonumber\\
{  N}_i{}^a&=&{\rm i}\,e^{2\,\varphi}\,\left[B_{ib}\,
E_4{}^b{}_{\hat{b}}\,E_4{}^a{}_{\hat{b}}+
(-B_{ib}\, C^{bc}+C_i^c)\,E_4^{-1}{}_c{}^{\hat{c}}\,E_4^{-1}{}_d{}^{\hat{c}}
\,C^{da}\right]+c^\prime\, C^a_i \,,
\nonumber\\
{  N}_{ia}&=&-{\rm i}\,e^{2\,\varphi}\,(-B_{ib}\, C^{bc}+C_i^c)\,
E_4^{-1}{}_c{}^{\hat{c}}\,E_4^{-1}{}_a{}^{\hat{c}}-c^\prime \,B_{ia} \,,
\nonumber\\
{  N}^{ij}&=&-{\rm i}\,e^{-2\,\varphi}\,E_2{}^i{}_{\hat{k}}\,
E_2{}^j{}_{\hat{k}} \,,
\nonumber\\
{  N}^{ia}&=& -\epsilon^{ij}\,C^a_j \,,
\nonumber\\
{  N}^{i}{}_a&=& \epsilon^{ij}\, B_{ja} \,,
\nonumber\\
{  N}^{ab}&=&-{\rm i}\,e^{2\,\varphi}\,\left(-C^{ad}\,
E_4^{-1}{}_d{}^{\hat{d}}\,E_4^{-1}{}_c{}^{\hat{d}}\,C^{cb}+ 
E_4{}^a{}_{\hat{b}}\,E_4{}^b{}_{\hat{b}}\right) \,,
\nonumber\\
{  N}^{a}{}_b&=&-{\rm i}\,e^{2\,\varphi}\,C^{ad}\,E_4^{-1}{}_d{}^{\hat{d}}\,
E_4^{-1}{}_b{}^{\hat{d}}+c^\prime\,\delta^a{}_b \,,
\nonumber\\
{  N}_{ab}&=&-{\rm
i}\,e^{2\,\varphi}\,E_4^{-1}{}_a{}^{\hat{d}}\,E_4^{-1}{}_b{}^{\hat{d}} \,.
\end{eqnarray}

\paragraph{Model $T_1\times T_5$}
The $Sp(24,\mathbb{R})$ representation of the $N_4$ generators is
the following:
\begin{eqnarray}
T^a &=&\left(\matrix{ 0&0&0&0&0&0&0&0\cr 0&0&0&0&0&0&0&0\cr
\delta^a_b&0&0&0&0&0&0&0\cr 0&\delta^a_b&0&0&0&0&0&0\cr 
0&0&0&0&0&0&-\delta^a_b&0\cr 0&0&0&0&0&0&0&-\delta^a_b\cr 0&0&0&0 &0&0&0&0\cr 
0&0&0 &0&0&0&0&0}\right) \,,
\nonumber\\
T_{a}&=&\left(\matrix{ 0&0&0&0&0&0&0&0\cr 0&0&0&0&0&0&0&0\cr
0&0&0&0&0&0&0&0\cr 0&0&0&0&0&0&0&0\cr 0&0&0&\delta^b_a&0&0&0&0\cr 
0&0&-\delta^b_a&0&0&0&0&0\cr 0&-\delta^b_a&0&0 &0&0&0&0\cr 
\delta^b_a&0&0 &0&0&0&0&0}\right) \,,
\nonumber\\
T_{ab}&=&\left(\matrix{ 0&0&0&0&0&0&0&0\cr 0&0&0&0&0&0&0&0\cr
0&0&0&0&0&0&0&0\cr 0&0&0&0 &0&0&0&0\cr 0&0&0&0&0&0&0 &0\cr 
0&0&0&0&0&0&0&0\cr 0&0&0&-\delta_{ab}^{cd}&0&0&0&0\cr 
0&0&\delta_{ab}^{cd} &0&0&0&0&0}\right)\nonumber\\
T&=&\left(\matrix{ 0&\bfone &0&0&0&0&0&0\cr 0&0&0&0&0&0&0&0\cr
0&0&0&\bfone &0&0&0&0\cr 0&0&0&0&0&0&0&0\cr 0&0 &0&0&0&0&0&0\cr 
0&0&0&0&-\bfone &0&0&0\cr 0&0&0&0&0&0&0&0\cr 0&0&0 &0&0&0&-\bfone&0}\right) \,.
\end{eqnarray}
We have chosen the coset representative to have the form given in eq. 
(\ref{coset1t5}). We may choose for the matrix $\mathbb{E}$ the following 
matrix form:
\begin{equation}
\mathbb{E} = \mbox{{\scriptsize $\left(\matrix{e^{\varphi}\,E &
0&0&0&0&0&0&0\cr 0&e^{-\varphi}\,E & 0&0&0&0&0&0\cr0&
0&e^{\varphi}\, E^{-1}{}_a{}^{\hat{b}} & 0&0&0&0&0\cr 0&0&0&
e^{-\varphi}\, E^{-1}{}_a{}^{\hat{b}} & 0&0&0&0\cr 0&0&0&0&
e^{-\varphi}/E&0&0&0\cr 0&0&0&0&0 &e^{\varphi}/E &0&0\cr 0&
0&0&0&0&0& e^{-\varphi}\, E^a{}_{\hat{b}} &0\cr 0& 0&0&0&0&0&0&
e^{\varphi}\, E^a{}_{\hat{b}} }\right)$}} \,,
\end{equation}
where:
\begin{eqnarray}
E^a{}_{\hat{b}}
\,,\,\, E&\in & O(1,1)_1\times O(1,1)_2\times
\frac{SL(5,\mathbb{R})}{SO(5)}\,,
\nonumber \\
e^{H\,\varphi} &\in & O(1,1)_0 \,.
\end{eqnarray}
The blocks  $\{{\sf A},\,{\sf C},\,{\sf D}\}$ of $L$ and ${\sf
A}^{-1}$ have the following form: 
{\small \begin{eqnarray} 
{\sf A}&=&\mbox{{\scriptsize $ \left(\matrix{E\, e^{\varphi}& c\,
E\,e^{-\varphi}&0&0\cr
0 & E\,e^{-\varphi} &0&0\cr B_a\, E\, e^{\varphi}& c\, B_a\, E\,
e^{-\varphi}& e^{\varphi}\, E^{-1}{}_a{}^{\hat{b}} & c\,e^{-\varphi} \,
 E^{-1}{}_a{}^{\hat{b}}\cr 0 &  B_a\, e^{-\varphi}&0 &e^{-\varphi} \,
  E^{-1}{}_a{}^{\hat{b}} }\right)$}} \,,
\nonumber\\
{\sf C}&=& \mbox{{\scriptsize $\left(\matrix{0& B_a\, C^a\,
E\,e^{-\varphi}&0& e^{-\varphi}\,(B_b\, C^{ba}+C^a)\,
E^{-1}{}_a{}^{\hat{d}} \cr -B_a\, C^a\, E\,e^{\varphi} & -c\,B_a\,
C^a\, E\,e^{-\varphi} &- e^{\varphi}\,(B_b\, C^{ba}+C^a)\,
E^{-1}{}_a{}^{\hat{d}}&- e^{-\varphi}\,c\,(B_b\, C^{ba}+C^a)\,
E^{-1}{}_a{}^{\hat{d}}\cr 0& -e^{-\varphi}\, C^a\, E&0&
e^{-\varphi}\, C^{ab}\, E^{-1}{}_b{}^{\hat{d}}\cr
C^a\, E\, e^{\varphi} &c\,C^a\, E\, e^{-\varphi}&e^{\varphi}\,C^{ab}\,
E^{-1}{}_b{}^{\hat{d}} &e^{-\varphi}\,c\,C^{ab}\, 
E^{-1}{}_b{}^{\hat{d}} }\right)$}} \,,
\nonumber\\
{\sf D}&=&\mbox{{\scriptsize $\left(\matrix{e^{-\varphi}/E & 0 & -B_a 
\,e^{-\varphi}\, E^a{}_{\hat{d}} & 0\cr - c\,e^{-\varphi}/E & e^{\varphi}/E 
& c\,B_a \,e^{-\varphi}\, E^a{}_{\hat{d}} & -B_a \,e^{\varphi}\, 
E^a{}_{\hat{d}} \cr 0 &0&e^{-\varphi}\, E^a{}_{\hat{d}} &0\cr  0 &0&-c\, 
e^{-\varphi}\, E^a{}_{\hat{d}}& e^{\varphi}\, E^a{}_{\hat{d}}}\right)$}} \,,
\nonumber\\
{\sf A}^{-1}&=& \mbox{{\scriptsize $\left(\matrix{e^{-\varphi}/E
&- c\,e^{-\varphi}/E  &0&0\cr 0 & e^{\varphi}/E & 0&0\cr-
e^{-\varphi}\, E_{\hat{d}}{}^a\, B_a & c\, e^{-\varphi}\,
E_{\hat{d}}{}^a\, B_a &  e^{-\varphi}\, E_{\hat{d}}{}^a &
-c\,e^{-\varphi}\, E_{\hat{d}}{}^a\cr 0 &-e^{\varphi}\,
E_{\hat{d}}{}^a B_a &0 & e^{\varphi}\, E_{\hat{d}}{}^a }\right)$}} \,,
\end{eqnarray}}
from equations (\ref{fh}) and (\ref{N}) we compute the matrix ${\Scr N}$:
\begin{equation}
{\Scr N} = \left(\matrix{N& N_1 & N^{(1a)}& N^a\cr *&
N_{1\,1}&N_1{}^{(1a)} & N_1{}^{a}
\cr *&*& N^{(1a)\,(1 b)} & N^{(1a)\, b}\cr *&*&*& N^{a\, b}} \right)
\,,
\end{equation}
whose entries are
\begin{eqnarray}
N&=& -{\rm i}\, e^{-2\,\varphi}\,(B_a\,B_b\, E^a{}_{\hat{a}}\,
E^b{}_{\hat{a}}+\frac{1}{E^2}) \,,
\nonumber\\
 N_1&=& {\rm i}\,c\,e^{-2\,\varphi}\,(B_a\,B_b\, E^a{}_{\hat{a}}\,
E^b{}_{\hat{a}}+\frac{1}{E^2}) \,,
\nonumber\\
N^{(1a)}&=& {\rm i}\,e^{-2\,\varphi}\,B_b\, E^a{}_{\hat{a}}\,
E^b{}_{\hat{a}} \,,
\nonumber\\
 N^a&=& C^a+B_b\, C^{ba}-{\rm i}\,c\,e^{-2\,\varphi}\,B_b\, 
E^a{}_{\hat{a}}\,E^b{}_{\hat{a}} \,,
\nonumber\\
N_{1\,1}&=& -{\rm i}\,(e^{-2\,\varphi}\, c^2+e^{2\,\varphi}) 
\,(B_a\,B_b\, E^a{}_{\hat{a}}\,E^b{}_{\hat{a}}+\frac{1}{E^2}) \,,
\nonumber\\
N_1{}^{(1a)}&=&-(B_b C^{ba}+C^a)-{\rm i}\,c\,e^{-2\,\varphi}\,B_b\, 
E^a{}_{\hat{a}}\,E^b{}_{\hat{a}} \,,
\nonumber\\
N_1{}^{a}&=& {\rm i}\,(e^{-2\,\varphi}\, c^2+e^{2\,\varphi})\,B_b\, 
E^a{}_{\hat{a}}\,E^b{}_{\hat{a}} \,,
\nonumber\\
 N^{(1a)\,( 1b)} &=&  -{\rm i}\,e^{-2\,\varphi}\,E^a{}_{\hat{a}}\,
E^b{}_{\hat{a}} \,,
\nonumber\\
N^{(1a)\, b}&=&-C^{ab}+{\rm i}\,e^{-2\,\varphi}\,c\,E^a{}_{\hat{a}}\,
E^b{}_{\hat{a}} \,,
\nonumber\\
N^{a\, b}&=&-{\rm i}\,(e^{-2\,\varphi}\, c^2+e^{2\,\varphi})\,
E^a{}_{\hat{a}}\,E^b{}_{\hat{a}} \,,
\end{eqnarray}
where the asterisks denote the symmetric entries.

\paragraph{Model $T_3\times T_3$}
The $Sp(24,\mathbb{R})$ representation of the $N_6$ generators is
the following:
\begin{eqnarray}
T_{ab} &=&\left(\matrix{ 0&0&0&0&0&0&0&0\cr 0&0&0&0&0&0&0&0\cr
0&0&0&0&0&0&0&0\cr 0&0&\delta_{ab}^{cd}&0&0&0&0&0\cr 0&0&0&0&0&0&0&0\cr 
0&0&0&0&0&0&0&0\cr 0&0&0&0 &0&0&0&\delta_{ab}^{cd}\cr 0&0&0 &0&0&0&0&0}
\right) \,,
\nonumber\\
T^{ia}&=&\left(\matrix{ 0&0&0&0&0&0&0&0\cr 0&0&0&0&0&0&0&0\cr
\delta^a_b\delta^i_j&0&0&0&0&0&0&0\cr 0&0&0&0&0&0&0&0\cr 
0&0&0&0&0&0&-\delta^i_j\delta^a_b&0\cr 0&0&0&\delta^i_j\delta^a_b&0&0&0&0\cr 
0&0&0&0 &0&0&0&0\cr 0&\delta^a_b\delta^i_j&0 &0&0&0&0&0}\right) \,,
\nonumber\\
T^{i}_a&=&\left(\matrix{ 0&0&0&0&0&0&0&0\cr 0&0&0&0&0&0&0&0\cr0&0&0&0&0&0&0&0
\cr
\delta^b_a\delta^i_j&0&0&0&0&0&0&0\cr 0&0&0&0&0&0&0&-\delta^i_j\delta^b_a\cr 
0&0&\delta^i_j\delta^b_a&0&0&0&0&0\cr 0&\delta^b_a\delta^i_j&0&0 &0&0&0&0\cr 
0&0&0 &0&0&0&0&0}\right) \,,
\nonumber\\
T^{ij}&=&\left(\matrix{ 0&0&0&0&0&0&0&0\cr 0&0&0&0&0&0&0&0\cr
0&0&0&0&0&0&0&0\cr 0&0&0&0 &0&0&0&0\cr 0&-2\,\delta^{ij}_{kl}&0&0&0&0&0 &0\cr
2\,\delta^{ij}_{kl}&0&0&0&0&0&0&0\cr 0&0&0&0&0&0&0&0\cr 0&0&0 &0&0&0&0&0}
\right) \,,
\nonumber\\
T&=&\left(\matrix{ 0&0 &0&0&0&0&0&0\cr \bfone&0&0&0&0&0&0&0\cr
0&0&0&0 &0&0&0&0\cr 0&0&0&0&0&0&0&0\cr 0&0 &0&0&0&-\bfone&0&0\cr 
0&0&0&0&0&0&0&0\cr 0&0&0&\bfone&0&0&0&0\cr 0&0&\bfone &0&0&0&0&0}\right) \,.
\end{eqnarray}
We have chosen the coset representative to have the form given in
eq. (\ref{coset3t3}). We may choose for the matrix $\mathbb{E}$
the following matrix form: 
\begin{equation}
\mathbb{E} = \mbox{{\scriptsize
$\left(\matrix{e^{-\varphi}\,E_1^i{}_{\hat{\jmath}} & 0&0&0&0&0&0&0\cr
0&e^{\varphi}\,E_1^i{}_{\hat{\jmath}}  & 0&0&0&0&0&0\cr0&
0&e^{-\varphi}\, E_2^{-1}{}_a{}^{\hat{b}} & 0&0&0&0&0\cr 0&0&0&
e^{-\varphi}\, E_2^a{}_{\hat{b}} & 0&0&0&0\cr 0&0&0&0&
e^{\varphi}\, E_1^{-1}{}_i{}^{\hat{\jmath}}&0&0&0\cr 0&0&0&0&0
&e^{-\varphi}\, E_1^{-1}{}_i{}^{\hat{\jmath}} &0&0\cr 0& 0&0&0&0&0&
e^{\varphi}\, E_2^a{}_{\hat{b}} &0\cr 0& 0&0&0&0&0&0&
e^{\varphi}\, E_2^{-1}{}_a{}^{\hat{b}} }\right)$}} \,,\nonumber
\end{equation}
where
\begin{eqnarray}
E_1^i{}_{\hat{\jmath}}&\in & \left(\frac{GL(3,\mathbb{R})}{SO(3)}\right)_1\,,
\nonumber \\
E_2^a{}_{\hat{b}} &\in & \left(\frac{GL(3,\mathbb{R})}{SO(3)}\right)_2\,,
\nonumber \\
e^{\varphi\, H} &\in & SO(1,1)_0 \,.
\end{eqnarray}
The blocks  $\{{\sf A},\,{\sf C},\,{\sf D}\}$ of $L$ and ${\sf A}^{-1}$ 
have the following form:
\begin{eqnarray}
{\sf A}&=& \mbox{{\scriptsize $\left(\matrix{ e^{-\varphi}\,
E_1^i{}_{\hat{\jmath}}&0&0&0\cr
c\, e^{-\varphi}\, E_1^i{}_{\hat{\jmath}}&e^{\varphi}\, 
E_1^i{}_{\hat{\jmath}} &0&0\cr B_{ia}\,
 e^{-\varphi}\,E_1^i{}_{\hat{\jmath}}& 0& e^{-\varphi}\, 
E^{-1}_2{}_a{}^{\hat{b}} &0\cr C_i^a\, e^{-\varphi}\, 
E_1^i{}_{\hat{\jmath}} &0 &e^{-\varphi} \,C^{ab}\, E^{-1}_2{}_b{}^{\hat{c}}&  
e^{-\varphi} \,E_2{}^a{}_{\hat{c}}}\right)$}} \,,
\nonumber\\
{\sf C}&=& \mbox{{\scriptsize $\left(\matrix{-2\,c\,e^{-\varphi}
\,\tilde{C}_{ij}\,E_1^j{}_{\hat{k}}&-2\,e^{\varphi}
\,\tilde{C}_{ij}\,E_1^j{}_{\hat{k}}&-c\,e^{-\varphi}
\,(C_i^a+B_{ib}\, C^{ba})\,E_2^{-1}{}_a{}^{\hat{c}}&
-e^{-\varphi}\,c\,B_{ib}\, E^{-1}_2{}^b{}_{\hat{d}} \cr
2 e^{-\varphi} \,\tilde{C}_{ij}\,E_1^j{}_{\hat{k}} & 0 &
e^{-\varphi} \,(C_i^a+B_{ib}\,
C^{ba})\,E_2^{-1}{}_a{}^{\hat{c}}&e^{-\varphi} \,B_{ia}\,
E_2{}^a{}_{\hat{c}} \cr c\,e^{-\varphi}\,C^a_i\,
E_1^i{}_{\hat{k}}& e^{\varphi}\,C^a_i\, E_1^i{}_{\hat{k}}&c\,
e^{-\varphi}\,C^{ab}\, E_2^{-1}{}_b{}^{\hat{c}}& c\,
e^{-\varphi}\,E_2{}^a{}_{\hat{c}}\cr
 c\, e^{-\varphi}\,B_{ia}\,E_1^i{}_{\hat{k}}&e^{\varphi}\,B_{ia}\,
E_1^i{}_{\hat{k}}&c\, e^{-\varphi}\,E_2^{-1}{}_a{}^{\hat{c}}&0}\right)$}} \,,
\nonumber\\
{\sf D}&=&\mbox{{\scriptsize $\left(\matrix{e^{\varphi}\,
E_1^{-1}{}_i{}^{\hat{\jmath}} & -c\,e^{-\varphi}\,
E_1^{-1}{}_i{}^{\hat{\jmath}} & -e^{\varphi}\,B_{ia} \, E_2^a{}_{\hat{d}} 
&-e^{\varphi} \,(C_i^a+B_{ib}\, C^{ba})\,E_2^{-1}{}_a{}^{\hat{c}}\cr 0 & 
e^{-\varphi}\,E_1^{-1}{}_i{}^{\hat{\jmath}} &0 & 0\cr 0 &0&e^{\varphi}\, 
E_2^a{}_{\hat{d}} &e^{\varphi}\,C^{ab}\,E_2^{-1}{}_b{}^{\hat{c}}\cr  
0 &0&0& e^{\varphi}\, E_2^{-1}{}_a{}^{\hat{c}}}\right)$}} \,,
\nonumber\\
{\sf A}^{-1}&=& \mbox{{\scriptsize $\left(\matrix{e^{\varphi}\,
E_1^{-1}{}^{\hat{\jmath}}{}_i &0 &0&0\cr -c\,e^{-\varphi}\,
E_1^{-1}{}^{\hat{\jmath}}{}_i& e^{-\varphi}\,E_1^{-1}{}^{\hat{\jmath}}{}_i 
& 0&0\cr- e^{\varphi}\,E_2{}_{\hat{a}}{}^a\,B_{ia} &0 &  e^{\varphi}\,
E_2{}_{\hat{a}}{}^a & 0\cr -e^{\varphi}\,E_2^{-1}{}^{\hat{c}}{}_a\,
(B_{ib}\, C^{ba}+C_i^a) &0& -e^{\varphi}\,E_2^{-1}{}^{\hat{c}}{}_a\, 
C^{ab} & e^{\varphi}\,E_2^{-1}{}^{\hat{c}}{}_a }\right)$}} \,,
\nonumber\\
\tilde{C}_{ij}&=&C_{ij}+{\textstyle \frac{1}{2}} \,C_i^a\, B_{ja} \,.
\end{eqnarray}
The vector kinetic matrix can now be calculated and has the following form:
\begin{equation}
{\Scr N} = \left(\matrix{N_{ij}& N^\prime_{ij} & N_i{}^a & N_{ia}\cr
*&N^{\prime\prime}_{ij}& N_i^\prime {}^a &  N_{ia}^\prime\cr *&*& 
N^{ab} &N^a{}_b\cr *&*&*& N_{ab} }\right) \,,
\end{equation}
whose entries are 
\begin{eqnarray}
N_{ij} &=& 2\, c\, B_{a(i}\, C^a_{j)}-{\rm i}\,\left[(c^2\,
e^{-2\,\varphi}+ e^{2\,\varphi})\,E_1^{-1}{}_i{}^{\hat{\jmath}}\,
E_1^{-1}{}_j{}^{\hat{\jmath}}+\right.
\nonumber\\
&& \left.e^{2\,\varphi}\,
(C_i^a+B_{ib}\, C^{ba})\,(C_j^c+B_{jb}\,
C^{bc})\,E_2^{-1}{}_a{}^{\hat{c}}\,E_2^{-1}{}_c{}^{\hat{c}}+
e^{2\,\varphi}\,B_{ia}\,B_{jb}\,E_2^a{}_{\hat{c}}\,
E_2^b{}_{\hat{c}}\right] \,,
\nonumber\\
N^\prime_{ij} &=&-2\,\tilde{C}_{ij}+{\rm
i}\,c\,e^{-2\,\varphi}\,E_1^{-1}{}_i{}^{\hat{\jmath}}\,
E_1^{-1}{}_j{}^{\hat{\jmath}} \,,
\nonumber\\
 N_i{}^a &=&-c\,C^a_i+{\rm i}\,e^{2\,\varphi}\,\left[(C_i^b+B_{ic}\, 
C^{cb})\,C^{ad}\,
 E_2^{-1}{}_b{}^{\hat{e}}\,E_2^{-1}{}_d{}^{\hat{e}}+B_{ib}\,E_2^a{}_{\hat{c}}\,
E_2^b{}_{\hat{c}}\right] \,,
\nonumber\\
N_{ia}&=& -c\,B_{ia}+{\rm i}\,e^{2\,\varphi}\,(C_i^b+B_{ic}\,
C^{cb})\,
 E_2^{-1}{}_b{}^{\hat{e}}\,E_2^{-1}{}_a{}^{\hat{e}} \,,
\nonumber\\
N^{\prime\prime}_{ij}&=&-{\rm
i}\,e^{-2\,\varphi}\,E_1^{-1}{}_i{}^{\hat{\jmath}}\,
E_1^{-1}{}_j{}^{\hat{\jmath}}
\nonumber\\
N_i^\prime {}^a &=& C_i^a \,,
\nonumber\\
N_{ia}^\prime &=& B_{ia} \,,
\nonumber\\
N^{ab}&=&-{\rm i}\,e^{2\,\varphi}\,\left[E_2^a{}_{\hat{c}}\,
E_2^b{}_{\hat{c}}+C^{a c}\,C^{bd}\,
E_2^{-1}{}_c{}^{\hat{e}}\,E_2^{-1}{}_d{}^{\hat{e}}\right] \,,
\nonumber\\
N^a{}_b&=&c-{\rm i}\,e^{2\,\varphi}\,C^{a c}\,
E_2^{-1}{}_c{}^{\hat{e}}\,E_2^{-1}{}_b{}^{\hat{e}} \,,
\nonumber\\
N_{ab}&=&-{\rm
i}\,e^{2\,\varphi}\,E_2^{-1}{}_a{}^{\hat{e}}\,E_2^{-1}{}_b{}^{\hat{e}} \,.
\end{eqnarray}

\paragraph{Model $T_5\times S_1$}
The $Sp(24,\mathbb{R})$ representation of the $N_8$ generators is
the following:
\begin{eqnarray}
T^i &=&\left(\matrix{ 0&0&0&0&0&0&0&0\cr
0&0&0&\delta^i_j&0&0&0&0\cr -\delta^i_j&0&0&0&0&0&0&0\cr
0&0&0&0&0&0&0&0\cr 0&0&0&0&0&0&\delta^i_j&0\cr 0&0&0&0&0&0&0&
0\cr 0&0&0&0 &0&0&0&0\cr 0&0&0 &0&0&-\delta^i_j&0&0}\right) \,,
\nonumber\\
T^{\prime i}&=&\left(\matrix{ 0&0&0&0&0&0&0&0\cr
0&0&-\delta^i_j&0&0&0&0&0\cr
0&0&0&0&0&0&0&0\cr \delta^i_j&0&0&0&0&0&0&0\cr 0&0&0&0&0&0&0&-\delta^i_j
\cr 0&0&0&0&0&0&0&0
\cr 0&0&0&0 &0&\delta^i_j&0&0\cr 0&0&0 &0&0&0&0&0}\right) \,,
\nonumber\\
T^{ij}&=&\left(\matrix{ 0&0&0&0&0&0&0&0\cr
\delta^{ij}_{kl}&0&0&0&0&0&0&0\cr
0&0&0&0&0&0&0&0\cr 0&0&0&0 &0&0&0&0\cr 0&0&0&0&0&\delta^{ij}_{kl}&0 &0\cr 
0&0&0&0&0&0&0&0
\cr 0&0&0&0&0&0&0&0\cr 0&0&0&0&0&0&0&0}\right) \,,
\nonumber\\
T&=&\left(\matrix{ 0&0 &0&0&0&0&0&0\cr 0&0&0&0&0&0&0&0\cr 0&0&0&0
&0&0&0&0\cr 0&0&0&0&0&0&0&0\cr 0&\bfone &0&0&0&0&0&0\cr
\bfone&0&0&0&0 &0&0&0\cr 0&0&0&\bfone &0&0&0&0\cr 0&0&\bfone
&0&0&0&0&0}\right) \,.
\end{eqnarray}
We have chosen the coset representative to have the form given in
eq. (\ref{coset5t1}). We may choose for the matrix $\mathbb{E}$
the following matrix form
{\small \begin{equation}
\mathbb{E} = \mbox{{\scriptsize
$\left(\matrix{e^{-\varphi}\,E^i{}_{\hat{\jmath}} & 0&0&0&0&0&0&0\cr
0&e^{-\varphi}\,E^{-1}{}_i{}^{\hat{\jmath}}  & 0&0&0&0&0&0\cr0&
0&e^{-\varphi}\, E & 0&0&0&0&0\cr 0&0&0& e^{-\varphi}\, /E &
0&0&0&0\cr 0&0&0&0& e^{\varphi}\, E^{-1}{}_i{}^{\hat{\jmath}}&0&0&0\cr
0&0&0&0&0 &e^{\varphi}\, E^i{}_{\hat{\jmath}} &0&0\cr 0& 0&0&0&0&0&
e^{\varphi}/E  &0\cr 0& 0&0&0&0&0&0& e^{\varphi}\, E
}\right)$}} \,, \nonumber
\end{equation}}
where:
\begin{eqnarray}
 E^i{}_{\hat{\jmath}},\,E&\in &
 O(1,1)_1\times O(1,1)_2\times\frac{SL(5,\mathbb{R})}{SO(5)}\,,
\nonumber \\
e^{\varphi\, H} &\in & SO(1,1)_0 \,.
\end{eqnarray}
The blocks  $\{{\sf A},\,{\sf C},\,{\sf D}\}$ of $L$ and ${\sf
A}^{-1}$ have the following form
\begin{eqnarray}
{\sf A}&=& e^{-\varphi}\,\mbox{{\scriptsize $\left(\matrix{
E^i{}_{\hat{\jmath}}&0&0&0\cr (B_{i}\,C_j+C_{ij})\, E^j{}_{\hat{\jmath}} &
E^{-1}{}_i{}^{\hat{\jmath}} &-B_{i}\,E&C_i/E\cr
-C_{i}\, E^i{}_{\hat{\jmath}}& 0&E &0\cr B_{i}\, E^i{}_{\hat{\jmath}}&0 
&0&  1/E}\right)$}} \,,
\nonumber\\
{\sf C}&=& e^{-\varphi}\,\mbox{{\scriptsize
$\left(\matrix{c\,(B_{i}\,C_j+C_{ij})\,E^j{}_{\hat{\jmath}}&
c\,E^{-1}{}_i{}^{\hat{\jmath}} &-c\,B_{i}\,E& c\,C_i/E \cr c\,
E^i{}_{\hat{\jmath}}& 0 &0&0 \cr c\,B_{i}\,E^i{}_{\hat{\jmath}}& 0&0& c/E
\cr
 -c\, C_i\,E^i{}_{\hat{k}}&0&c\, E&0}\right)$}} \,,
\nonumber\\
{\sf D}&=& e^{\varphi}\,\mbox{{\scriptsize
$\left(\matrix{E^{-1}{}_i{}^{\hat{\jmath}} &
(B_i\,C_j+C_{ij})\,E^j{}_{\hat{\jmath}} &C_i/E &-B_i \,E\cr 0 &
E^i{}_{\hat{\jmath}} &0 & 0\cr 0 &B_i E^i{}_{\hat{\jmath}}& 1/E & 0\cr  0
&-C_i E^i{}_{\hat{\jmath}}&0& E}\right)$}} \,,
\nonumber\\
{\sf A}^{-1}&=& e^{\varphi}\,\mbox{{\scriptsize
$\left(\matrix{E^{-1}{}^{\hat{\jmath}}{}_i &0 &0&0\cr
E_{\hat{\imath}}{}^j\,(C_j\,B_i-C_{ji})& E_{\hat{\imath}}{}^j&
E_{\hat{\imath}}{}^j\,B_j&-E_{\hat{\imath}}{}^j\,C_j\cr C_i/E&0 &
 1/E & 0\cr
 -B_i\,E&0& 0 &E }\right)$}} \,.\nonumber
\end{eqnarray}
The vector kinetic matrix can now be calculated and has the
following form
\begin{equation}
{\Scr N} = \left(\matrix{N_{ij}& N_{i}{}^j & N_i &
N_{i}^\prime\cr
*&N^{ij}& N^i &  N^{\prime i}\cr *&*& N &N^\prime \cr *&*&*& 
N^{\prime\prime} }\right) \,,
\end{equation}
whose entries are
\begin{eqnarray}
N_{ij}&=& -{\rm
i}\,e^{2\,\varphi}\,\left[E^{-1}{}_i{}^{\hat{\jmath}}\,
E^{-1}{}_j{}^{\hat{\jmath}}+(B_i\,
C_k+C_{ik})\,(B_j\,
C_n+C_{jn})\,E^k{}_{\hat{\ell}}\,E^n{}_{\hat{\ell}}+
\frac{1}{E^2}\,C_i\,C_j+B_i\,
B_j\, E^2\right] \,,
\nonumber\\
N_{i}{}^j&=&c-{\rm i}\,e^{2\,\varphi}\,\left[(B_i\, C_k+C_{ik})\,
E^k{}_{\hat{\ell}}\,E^j{}_{\hat{\ell}}\right] \,,
\nonumber\\
N_i&=&-{\rm i}\,e^{2\,\varphi}\,\left[(B_i\,
C_k+C_{ik})\,E^k{}_{\hat{\ell}}\,
B_j\,E^j{}_{\hat{\ell}}+\frac{1}{E^2}\,C_i\right] \,,
\nonumber\\
N_{i}^\prime &=&{\rm i}\,e^{2\,\varphi}\,\left[(B_i\, C_k+C_{ik})\,
E^k{}_{\hat{\ell}}\,
C_j\,E^j{}_{\hat{\ell}}+E^2\,B_i\right] \,,
\nonumber\\
N^{ij}&=&-{\rm i}\,e^{2\,\varphi}\,E^i{}_{\hat{\ell}}\,E^j{}_{\hat{\ell}} \,,
\nonumber\\
N^i &=& -{\rm i}\,e^{2\,\varphi}\,B_j\,E^i{}_{\hat{\ell}}\,
E^j{}_{\hat{\ell}} \,,
\nonumber\\
N^{\prime i}&=&{\rm i}\,e^{2\,\varphi}\,C_j\,E^i{}_{\hat{\ell}}\,
E^j{}_{\hat{\ell}} \,,
\nonumber\\
N&=&-{\rm i}\,e^{2\,\varphi}\,\left[B_i\,B_j\,E^i{}_{\hat{\ell}}\,
E^j{}_{\hat{\ell}}+\frac{1}{E^2}\right] \,,
\nonumber\\
N^\prime &=& c+{\rm i}\,e^{2\,\varphi}\,C_i\,B_j\,
E^i{}_{\hat{\ell}}\,E^j{}_{\hat{\ell}} \,,
\nonumber\\
N^{\prime\prime}&=&-{\rm
i}\,e^{2\,\varphi}\,\left[C_i\,C_j\,E^i{}_{\hat{\ell}}\,
E^j{}_{\hat{\ell}}+E^2\right] \,.
\end{eqnarray}

\end{document}